 \documentclass[twocolumn]{aastex63}  

\shorttitle{NGC 4424 \& Nikhuli}
\shortauthors{Graham et al.}

\begin{document}

\title{Potential Black Hole Seeding of the Spiral Galaxy NGC~4424 via an Infalling Star Cluster} 

\author[0000-0002-6496-9414]{Alister W.\ Graham}
\affiliation{Centre for Astrophysics and Supercomputing, Swinburne University of
  Technology, Victoria 3122, Australia}
\email{AGraham@swin.edu.au}

\author[0000-0002-4622-796X]{Roberto Soria}
\affiliation{College of Astronomy and Space Sciences, University of the Chinese
  Academy of Sciences, Beijing 100049, China} 
\affiliation{Sydney Institute for Astronomy, School of Physics A28, The University of Sydney, Sydney, NSW 2006, Australia}

\author{Bogdan C.\ Ciambur}
\affiliation{LERMA, Observatoire de Paris, CNRS, PSL University, Sorbonne Universités, UPMC
Univ.\ Paris 06, F-75014 Paris, France}
\affiliation{GEPI, Observatoire de Paris, PSL Research University, Place Jules Janssen,
F-92190 Meudon, France}

\author[0000-0002-4306-5950]{Benjamin L.\ Davis}
\affiliation{Centre for Astrophysics and Supercomputing, Swinburne University of
  Technology, Victoria 3122, Australia}
\affiliation{Center for Astro, Particle, and Planetary Physics (CAP$^3$), New
  York University Abu Dhabi, United Arab Emirates}

\author{Douglas A.\ Swartz}  
\affiliation{Astrophysics Office, NASA Marshall Space Flight Center, ZP12, Huntsville, AL 35812, USA}

\begin{abstract}

Galaxies can grow through their mutual gravitational attraction and subsequent
union.  While orbiting a regular high-surface-brightness galaxy, the body of a
low-mass galaxy can be stripped away. However, the stellar heart of the
infalling galaxy, if represented by a tightly bound nuclear star cluster, is
more resilient.  From archival Hubble Space Telescope images, we have
discovered a red, tidally-stretched star cluster positioned $\sim$5$\arcsec$ 
($\sim$400~pc in projection) from, and pointing toward the center of, the
post-merger spiral galaxy NGC~4424.  The star cluster, which we refer to as
`Nikhuli', has a near-infrared luminosity of
$(6.88\pm1.85)\times10^6\,$L$_{\odot,F160W}$ and likely represents the
nucleus of a captured/wedded galaxy.  Moreover, from our Chandra X-ray
  Observatory image, Nikhuli is seen to contain a high-energy X-ray point
source, with $L_{0.5-8\,{\rm keV}}=6.31^{+7.50}_{-3.77}\times10^{38}$ erg
s$^{-1}$ (90\% confidence). We argue that this is more likely to be an active
massive black hole than an X-ray binary.  Lacking an outward-pointing
comet-like appearance, the stellar structure of Nikhuli favors infall rather
than the ejection from a gravitational-wave recoil event.  A minor merger with
a low-mass early-type galaxy may have sown a massive black hole, aided an
X-shaped pseudobulge, and be sewing a small bulge.  The stellar mass and the
velocity dispersion of NGC~4424 predict a central black hole of
(0.6--1.0)$\times10^5\,M_{\odot}$, similar to the expected intermediate-mass
black hole in Nikhuli, and suggestive of a black hole supply mechanism for
bulgeless late-type galaxies.  We may potentially be witnessing black hole
seeding by capture and sinking, with a nuclear star cluster the delivery vehicle. 

\end{abstract}


\section{Introduction}

Galaxies have long been known to merge. 
Smaller galaxies rain down on bigger galaxies, stoking their growth 
\citep{1927MNRAS..87..420L, 1951ApJ...113..413S,
  1954ApJ...119..206B, 1959VV....C......0V, 2019ApJ...872..152I}.  This
evolution has
become a defining tenet of the cold dark matter model.  The ongoing discovery of
(infalling) satellites continues to test and support the hierarchical growth
scenario for galaxies.  The burgeoning detection of drawn-out, low surface brightness,
debris streams of former galaxies in the outskirts of larger galaxies
\citep{1998Natur.394..752P, 2006ApJ...642L.137B, 2014ApJ...787...19M, 
2015ApJ...813..109D, 2015MNRAS.446.3110K, 2018ApJ...862..114S,
  martinezdelgado2021hidden} are starting to reveal the extent of the
hierarchical model, at least in regard to building stellar halos.  The closest known
tidal stream to the center of our galaxy is 10-20 kpc distant from the galaxy
center \citep{2020ApJ...898L..37Y, 2021ApJ...920...51M}. 

The dense 
and compact nuclear star clusters---which are  common features at the
centers 
of low ($\approx$10$^7$--10$^9\,M_{\odot}$) and intermediate stellar mass 
($\approx$$10^9$--$10^{10.5}\,M_{\odot}$) 
galaxies\citep[e.g.][]{1985AJ.....90.1759S,
  2002AJ....123.1389B, 2003AJ....125.2936G, 2003ApJ...582L..79B,
  2007ApJ...665.1084B, 2013ApJ...763...76S, 2019ApJ...878...18S}---are 
more strongly gravitationally bound than their
host galaxies and are thus more able to resist disruption during the
cannibalistic process 
\citep[e.g.,][]{2001ApJ...552L.105B, 2003MNRAS.344..399B}.  Indeed, the dual
and multiple nuclei of massive early-type galaxies is a testament to the
robustness of these captured star clusters during the galaxy assimilation
process 
\citep[e.g.,][]{1974ApJ...191...55J, 1978Afz....14...69P, 1981Afz....17..231K,
  1985AJ.....90.2431T, 1992AJ....103.1457B, 2016ApJ...829...81B}.  
Furthermore, the abundance of kinematically distinct cores and nuclear stellar disks
in early-type galaxies  
\citep[e.g.][]{1989ApJ...344..613F, 1995AJ....109.1988F, 1998A&AS..133..325G, 1999MNRAS.306..437H} 
reveals a more complete journey to the center of these galaxies, in stark contrast
with an eternity spent in the halo. 

Such galaxy growth, revealed by 
off-center nuclei, can be harder to spot in late-type galaxies due 
to confusion with knots of star formation. 
This ambiguity makes the later stages of the merger/digestion process more
challenging 
to establish in late-type galaxies. Furthermore, 
without a massive stellar bulge, dynamical friction will be less
in the inner regions, 
reducing the extent to which minor mergers contribute to 
the nuclei and bulges. 
The abundance of low bulge-to-total stellar mass ratios in late-type galaxies identified by
\citet{2008MNRAS.388.1708G} and \citet{2009ApJ...696..411W} was later 
used by \citet{2010ApJ...723...54K} to challenge the hierarchical growth of
most late-type galaxies.  It was speculated that the 
bulges were too small to be built by
mergers and were thus considered the result of internal secular processes. 
The detection of an infalling nuclear star cluster near the center of a late-type
spiral galaxy would reveal that the Roche-lobe destruction of 
galactic neighbors does not just produce streams that build diffuse
halos, but that tidal-shredding also contributes to the 
galaxy proper, including small bulges. 
Moreover, such digestion could potentially seed and feed spiral galaxies with
massive black holes. 

Among the Virgo cluster's spiral galaxies, the `Sa? pec'
\citep{1981rsac.book.....S} galaxy NGC~4424 (VCC
979), with a disk-inclination of 70$\deg$ from face-on, 
is of particular interest.  While there is no detectable X-ray point
source at the center of NGC~4424, there is a nearby ($\sim$5$\arcsec$)
`off-center' source, NGC~4424 X-3 ($L_{0.5-8\,{\rm keV}} \approx 6\times
10^{38}$ erg s$^{-1}$).  First publicly reported by
\citet{2018A&A...620A.164B}, they suggested it is either an X-ray binary (XRB)
or the nucleus of a merged galaxy.  Two brighter X-ray point sources plus a
fainter fourth point source reside in the outskirts of this galaxy.

Although \citet{2011NewAR..55..166F}, \citet{2014Natur.514..202B} and
\citet{2017ARA&A..55..303K} reveal why super-Eddington accretion
\citep{2014Sci...345.1330A, 2014Sci...343.1330S} is thought to explain many of
the off-center ultraluminous X-ray sources (ULX: $L_{\rm X} \approx 3\times 10^{39}$
to $10^{41}$ erg s$^{-1}$), some might be 
intermediate-mass black holes (IMBHs) from accreted galaxies,
particularly the hyperluminous X-ray sources \citep[HLXs: $L_{\rm X} > 10^{41}$ erg
  s$^{-1}$,][]{2003ApJ...596L.171G} such as ESO~243-49
\citep{2009Natur.460...73F, 2010MNRAS.405..870S} and others
\citep{2012MNRAS.423.1154S, 2019ApJ...882..181B, 2019MNRAS.483.5554E}.  Here,
we report the association of NGC~4424 X-3 with a red, elongated star cluster.

In Section~\ref{Sec_Data}, we provide some background information and
references to NGC~4424 before presenting optical and near-infrared images
along with an X-ray map.  We perform a re-analysis of the Hubble Space
  Telescope (HST) image, the Chandra X-ray Observatory (CXO)
and XMM-Newton data, and provide an extended discussion.  Section~\ref{Sec_BH}
offers a prediction for the mass of the central black hole in both NGC~4424
and the suspected infaller.  Finally, a more detailed discussion, centered
around off-centered sources, is provided in Section~\ref{Sec_Disc}.

Previously, in \citet{2019MNRAS.484..814G}, we used 
the mean redshift-independent distance modulus for NGC~4424 of 30.6$\pm$1.0 
from the  NASA/IPAC Extragalactic Database
(NED)\footnote{http://nedwww.ipac.caltech.edu}. 
Here, we adopt the Cepheid-based distance modulus of
31.080$\pm$0.292 (luminosity distance $D=16.44^{+2.37}_{-2.07}$~Mpc) from 
\citet{2016ApJ...826...56R}, giving a scale of 79 pc arcsec${-1}$.  
This encompasses and agrees with an array of recent results, including 
\citet{2018ApJ...861..104H}, who report a `tip of the red giant 
branch' distance modulus of $31.00 \pm 0.03_{\rm stat} \pm 0.06_{\rm sys}$ 
($15.8\pm0.2_{\rm stat}\pm0.4_{\rm sys}$ Mpc), and with
\citet{2013NewA...20...30M}, who derived a SN Ia distance modulus of 30.95, and
with \citet{2008ApJ...683...78C}, who reported a value of 30.91 based on the
\citet{1977A&A....54..661T} relation.

\section{Observations and Findings} \label{Sec_Data}

\subsection{Preface} 

From deep {\it CXO} exposures of 100
early-type galaxies in the Virgo cluster---half of which possess a nuclear star
cluster \citep{2006ApJS..165...57C,
  2006ApJS..164..334F}---\citet{2010ApJ...714...25G}  
effectively reveals that nuclear star clusters, in
and of themselves, tend not to contain X-ray point sources.  Among the 
30 lowest mass galaxies of that 100-strong sample,  
most are nucleated, and all have been predicted to harbour an
IMBH.  \citet{2019MNRAS.484..794G} reported that 
just three of these 30 contained a central X-ray point source, ruling out stellar-mass
XRBs as a common feature of nuclear star clusters, at least in early-type galaxies. 
If these star clusters are comprised of both a young and old stellar population, it
will rule out both high- and low-mass XRBs, respectively. 

\begin{figure}
\includegraphics[angle=00,width=1.0\columnwidth]{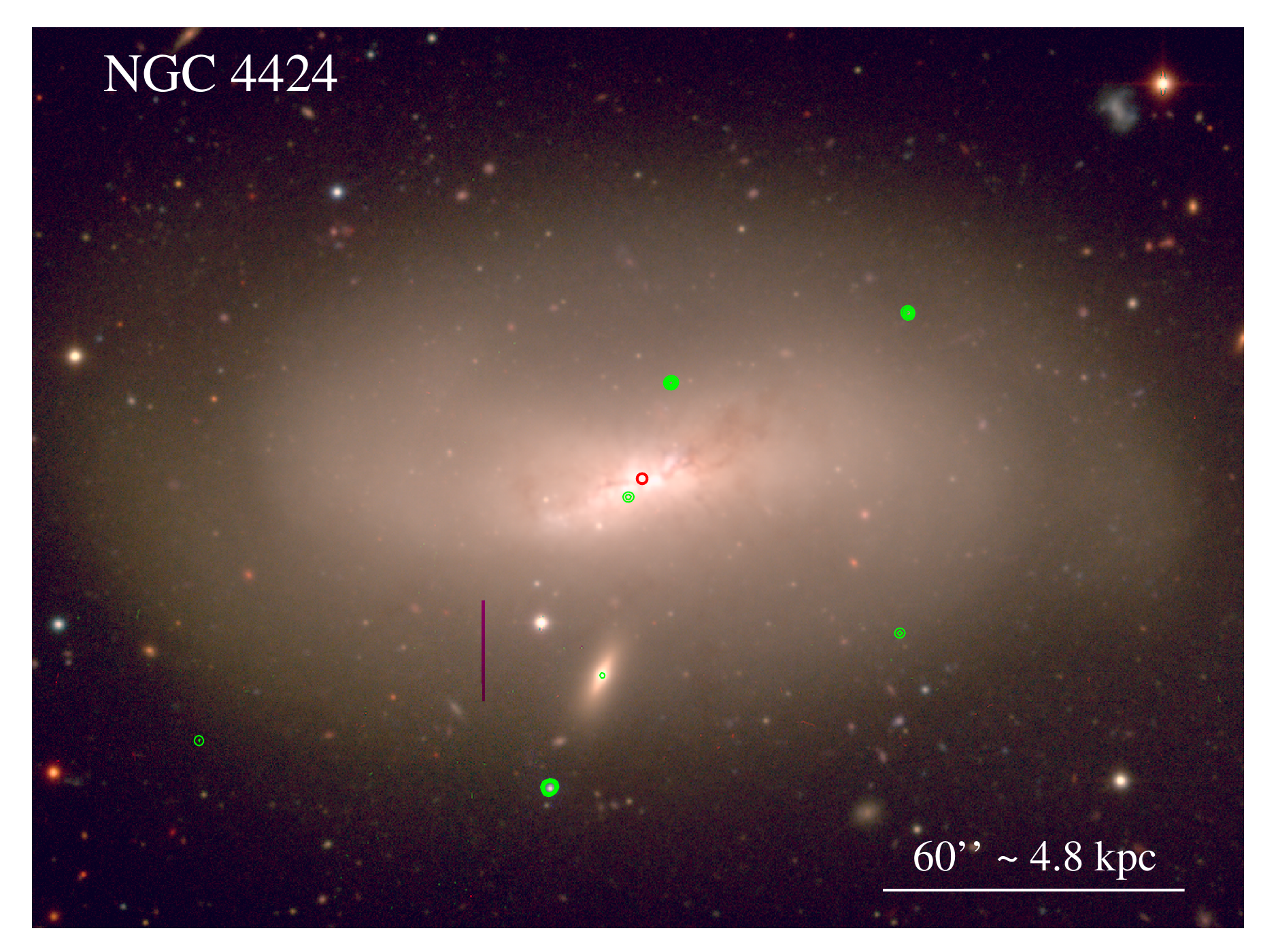}
  \caption{Next Generation Virgo Cluster Survey
    \citep[NGVS:][]{2012ApJS..200....4F} image of the post-merger galaxy
    NGC~4424 (red = $i$ filter; green = $g$; blue = $u$), with {\it CXO}/ACIS-S contours
    (0.5--7.0 keV band) overlaid in green. North is up, east is to the
    left. The red circle shows the NED-provided position for the galaxy's
    optical nucleus, with a radius of 1 arcsec capturing the associated
    uncertainty.  The near-central, but off-center, X-ray point source
    (NGC~4424 X-3) is associated with an elongated star cluster 
    (see Figures~\ref{Fig-HLA-core} and \ref{Fig_4424_resid}).  
    The galaxy to the south
    is IC~3366 (LEDA~213994), a background galaxy at a distance of $\sim$110 Mpc, while the
    vertical red stripe is an artifact that should be ignored.}
\label{Fig-NGVS}
\end{figure}

From past observations, 
NGC~4424 (Figure~\ref{Fig-NGVS}) has been identified as a post-merger
galaxy experiencing central star formation
\citep{1996AJ....111..152K}.  
\citet{2006AJ....131..747C} and \citet{2018A&A...620A.164B} provide a detailed
analysis of NGC~4424, finding that it experienced an unequal-mass
merger event less than 0.5~Gyr ago. The encounter has fueled an outburst of
ionized gas, evident as a $\sim$10~kpc plume pointing in the opposite
direction to a $\sim$110~kpc long (ram pressure)-stripped tail of
H{\footnotesize I} gas emanating from NGC~4424 as it 
plows through the hot intracluster medium of the Virgo cluster.

\begin{figure}
\includegraphics[angle=00,width=0.95\columnwidth]{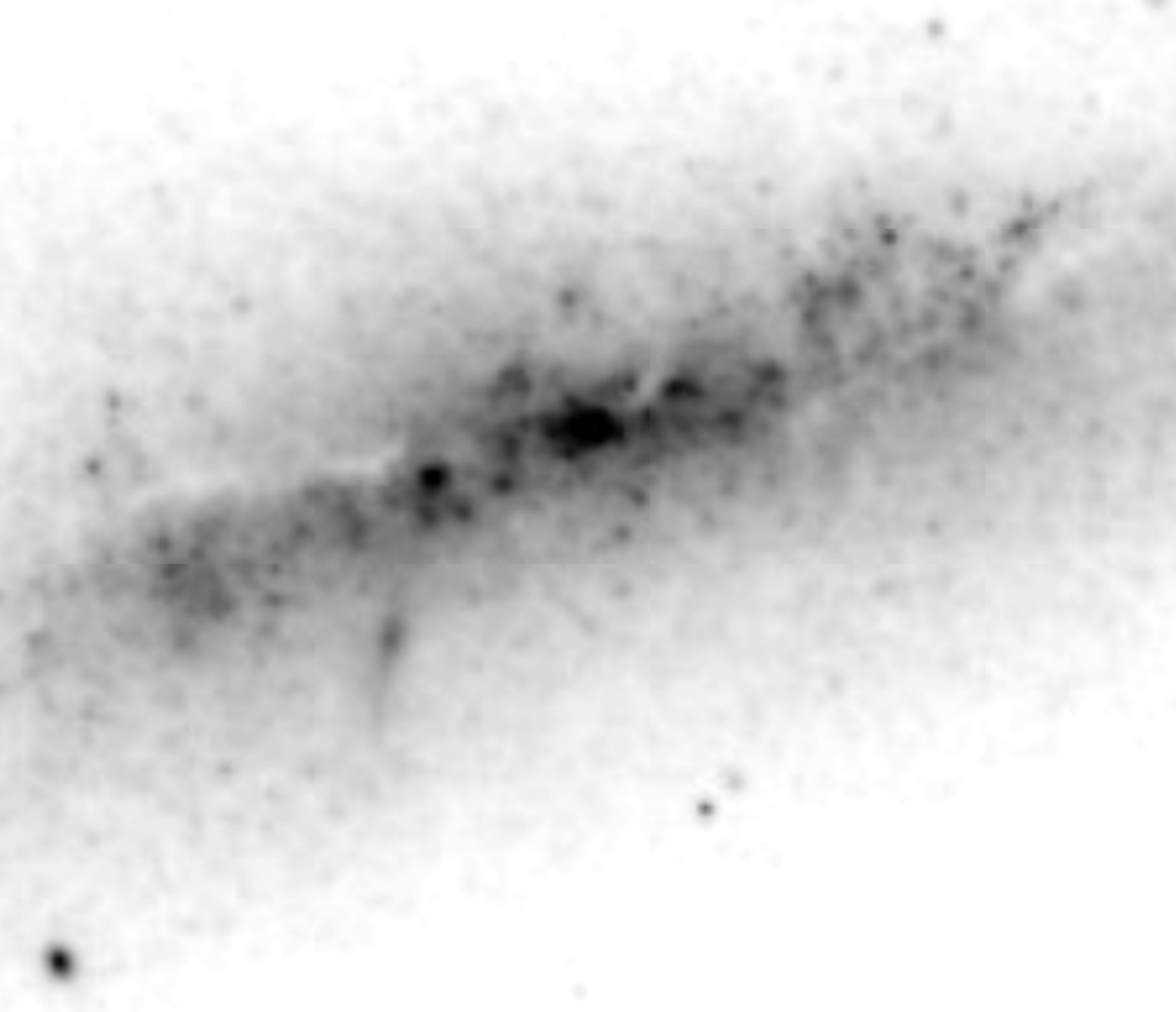}
  \caption{{\it HST}/WFC3-IR F160W image of the inner 
    $21\arcsec\times18\arcsec$ region of NGC~4424, 
    displaying an elongated arc or stream with a nucleus 5$\arcsec$ to the southeast of the
    galaxy center.  The near-central X-ray point-source in
    Figure~\ref{Fig-NGVS} coincides with this nucleus. 
   North is up, east is to the left.  
  Image: HLA Dataset hst\_12880\_26\_wfc3\_ir\_f160w.}
\label{Fig-HLA-core}
\end{figure}

\begin{figure*}
\includegraphics[angle=00,width=1.0\textwidth]{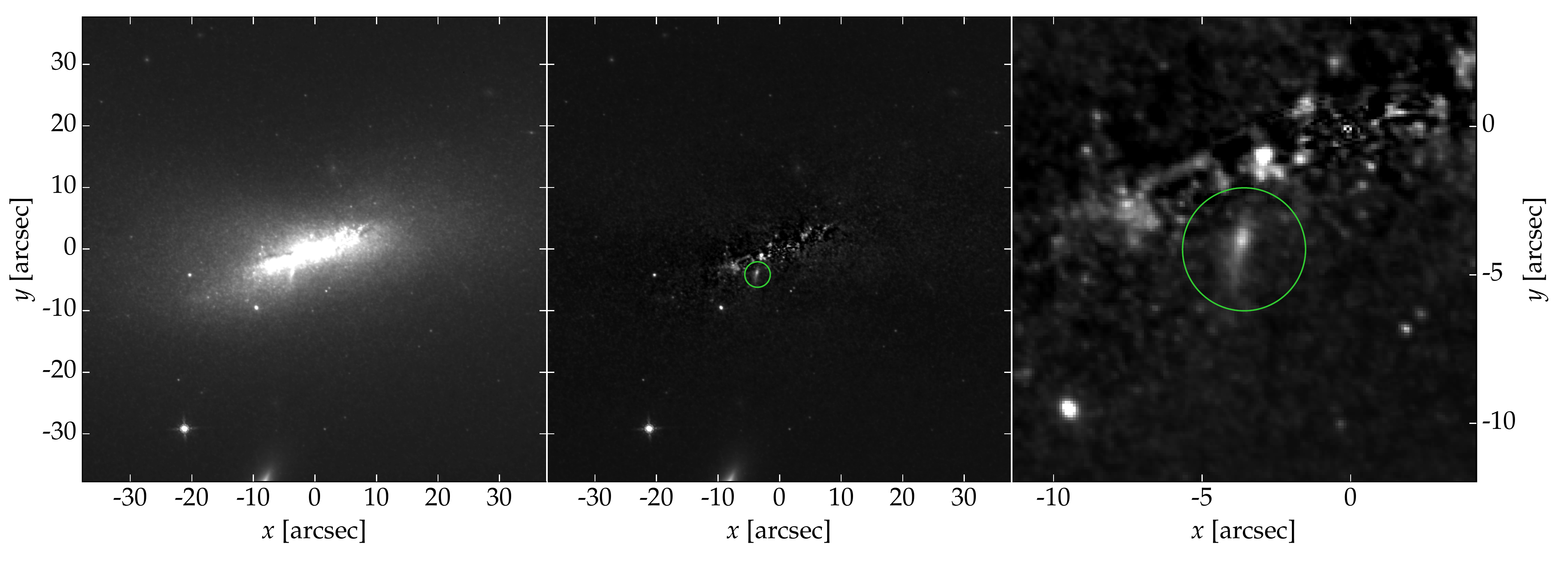}
  \caption{NGC~4424: {\it HST}/WFC3-IR F160W cut-out image (left panel:
    $88\arcsec \times 88\arcsec \approx 7 {\rm kpc} \times 7 {\rm kpc}$); 
    {\it Isofit} residual image
    (middle panel); zoom-in of the region containing NGC~4424 X-3 (right panel).
}
\label{Fig_4424_resid}
\end{figure*}

\subsection{Optical/near-IR images} 
\label{Sec_Opt}


From Figure~\ref{Fig-NGVS}, obtained as a part of the Next Generation Virgo
Cluster Survey \citep[NGVS:][]{2012ApJS..200....4F}, NGC~4424 appears as a
bulgeless, or near-bulgeless, galaxy containing a knotty bar-like feature with
an ellipticity to great to be a (relaxed) bulge.  While
\citet{1991rc3..book.....D} catalog this galaxy as having spiral arms and a
bar, this bar-like feature does not resemble a typical (relaxed) bar.
Instead, it displays an apparent X/(peanut shell)-shaped (X/P) structure; a
pseudobulge formed from the buckling of a bar
\citep{1975IAUS...69..297B,1975IAUS...69..349H,1981A&A....96..164C,1990A&A...233...82C,2005MNRAS.358.1477A,2018ApJ...852..133S}.
The broad, faint spiral arms emanate from the ends of the X-shaped bar-like
feature, at a major-axis radius of $\sim$36$\arcsec$.

Zooming in on the central region, and dialing down the brightness, 
Figure~\ref{Fig-HLA-core} shows part of an image acquired from the 
Hubble Legacy Archive (HLA)\footnote{\url{https://hla.stsci.edu/}}.
It displays the inner region of a Wide Field Camera 
3 (WFC3) IR channel F160W ($H$-band) image 
taken as a part of {\it HST} Proposal ID 12880
(PI: A.Riess).  
The near-IR exposure helps to reduce the impact of dust, young stars and avoid H$\alpha$ emission. 
One can see a nucleated stream to the lower-left (southeast) of the galaxy center. 

Figure~\ref{Fig_4424_resid} shows our modeling and removal of
much of the galaxy's mirror-symmetric light about its major-axis (having a variable 
position angle with radius).  
Our data reduction process followed that detailed in \citet{2019ApJ...873...85D}.
We modeled NGC~4424 using the modified task {\it Isofit} 
\citep{2015ApJ...810..120C}, run within the Image Reduction and Analysis
Facility ({\sc IRAF})\footnote{\url{http://ast.noao.edu/data/software}} package {\sc
  ellipse}.  Our model consists of a series of one-dimensional profiles
including the radial run of intensity, major-axis position angle, ellipticity
and a sequence of higher-order Fourier harmonic terms capturing the isophotes'
departures from pure ellipses.  

Our two-dimensional galaxy model --- built from the assorted one-dimensional
profiles using the new {\it Cmodel} task \citep{2015ApJ...810..120C} within
{\sc IRAF} 
--- has been subtracted from the original image to highlight the
residual structure (Figure~\ref{Fig_4424_resid}). 
This better reveals the elongated, off-center 
star cluster $\sim$5$\arcsec$ ($\sim$400 pc) to the southeast of the galaxy
center.  We have subsequently discovered that this previously overlooked structure is also visible in 
the F160W/F814W/F555W composite image made by  Lisa Frattare, which can be seen at NASA's
website\footnote{\url{https://www.nasa.gov/image-feature/goddard/2017/hubbles-double-galaxy-gaze-leda-and-ngc-4424}}
and in \citet[][their Figure~1]{2016ApJ...819...31G}. 
Moreover, this star cluster is also where the off-centered X-ray point source resides 
\citep[Figure~\ref{Fig-NGVS}, see also ][]{2018A&A...620A.164B}. 
This find, or realization, is reminiscent of the discovery of the 
optical counterpart/host of HLX-1 in the galaxy ESO~243-49
\citep{2010MNRAS.405..870S}, which can be clearly seen in Figure~7 of 
\citep{2015ApJ...810..120C}.  

\begin{figure*}
$
\begin{array}{cc}
 \includegraphics[angle=0, trim=0cm 0.0cm 0cm 0cm,
   width=1.0\columnwidth]{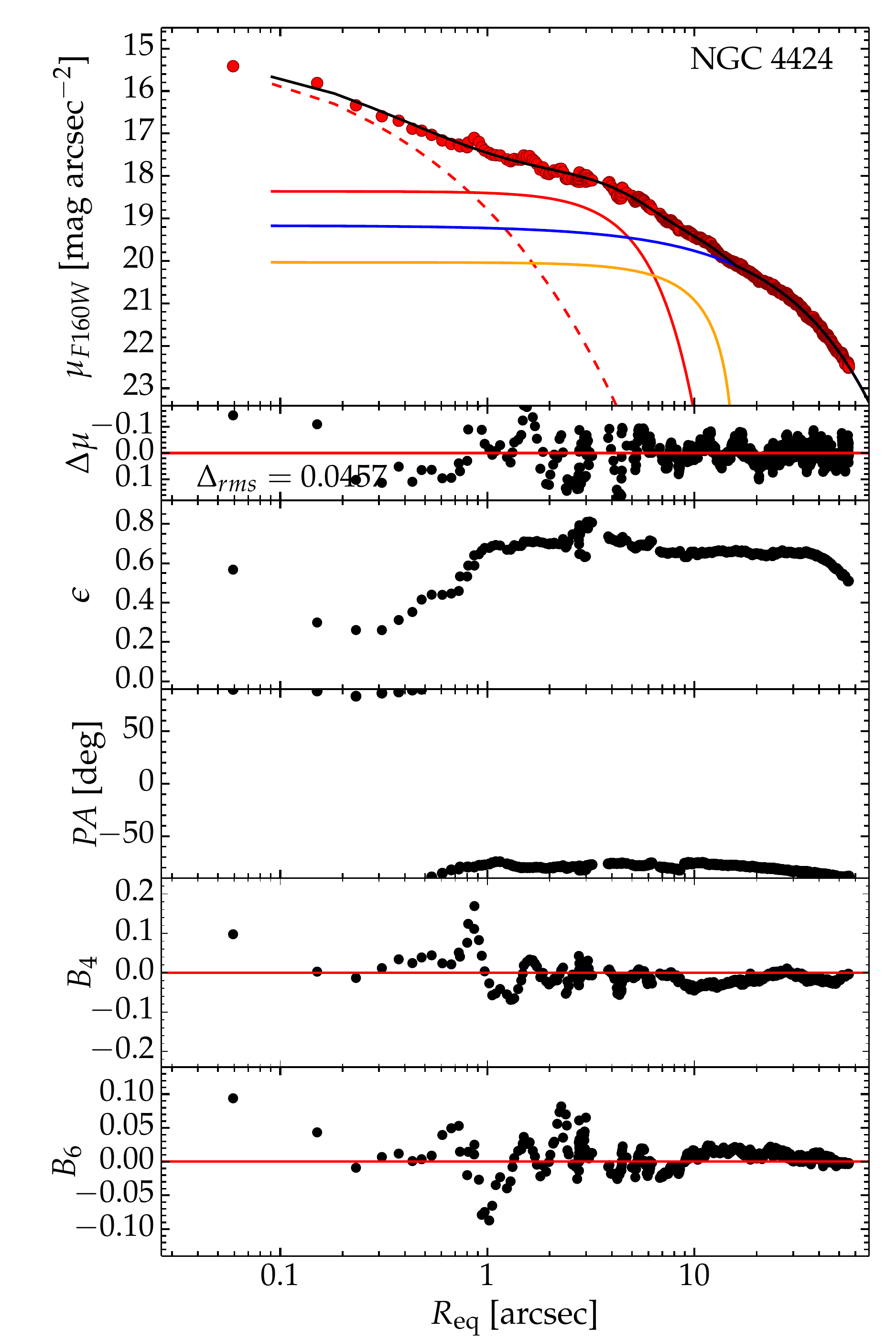} &
 \includegraphics[angle=0, trim=0cm 0.0cm 0cm 0cm,
   width=1.0\columnwidth]{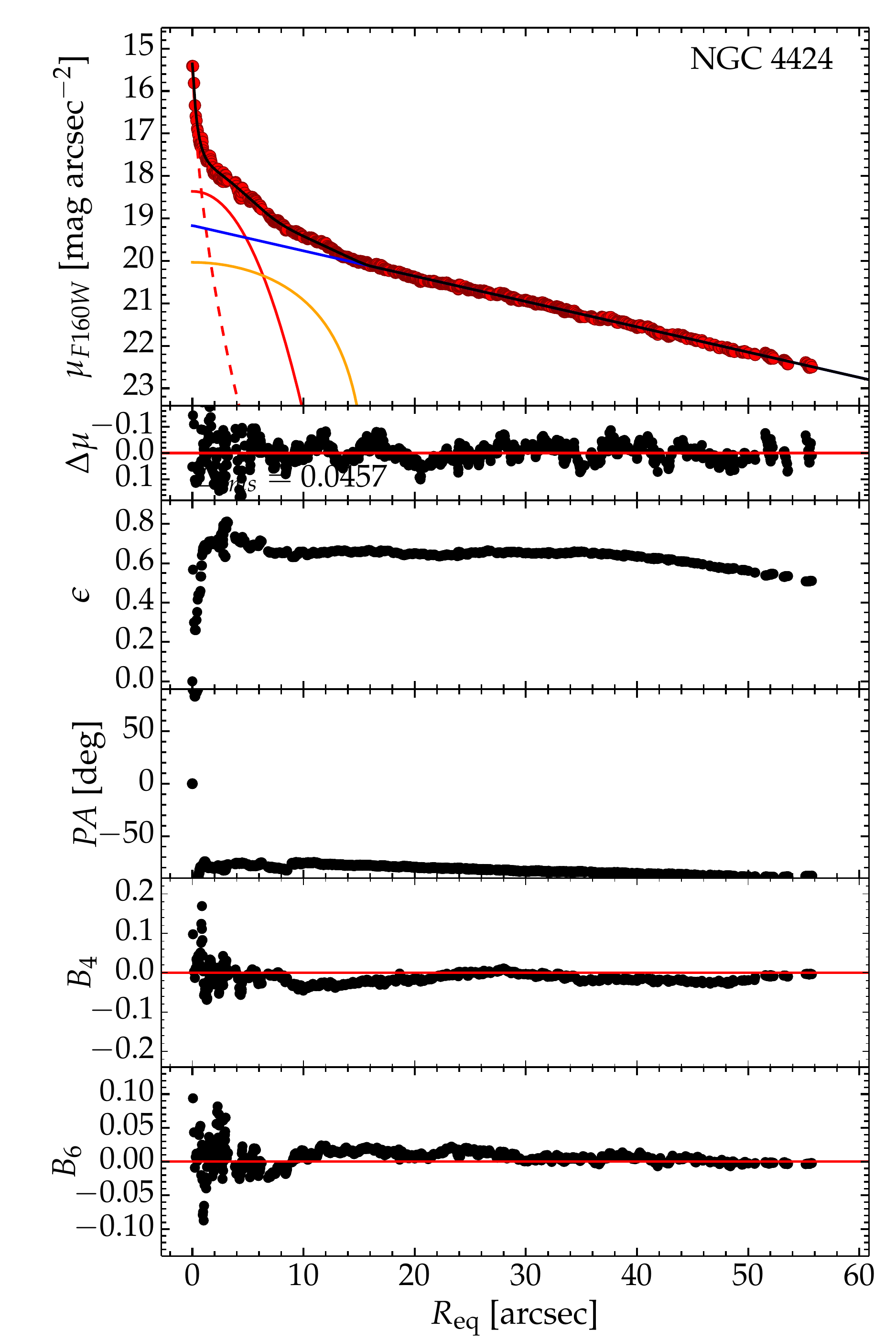} 
 \\
\end{array}
$
\caption{Geometric-mean axis, aka equivalent axis, of the 
{\it HST}/WFC3-IR F160W ($H$-band) light profile of NGC~4424.  The
  galaxy light (red points) has been decomposed into structures using the {\sc
    Profiler} software \citep{2016PASA...33...62C}.  While there is clearly a
  large-scale disk (dark blue), the unrelaxed nature of this post-merger
  remnant makes the inner components somewhat speculative.  We have used a
  Ferrer bar component (orange), a S\'ersic pseudobulge component (solid red), plus a
  Gaussian for the central star clusters (dashed red).} 
\label{Fig-Prof}
\end{figure*}

NGC~4424 contains a well-known, large-scale exponential-disk 
\citep{1996AJ....111..152K,2006AJ....131..747C}. 
Within this disk is a long faint-bar whose inner half has experienced a
brightening, associated the boxy X/P structure or pseudobulge. 
The negative $B_4$ term at $\sim$12 arcseconds (Figure~\ref{Fig-Prof}) 
reflects the boxy isophotes of the bar/pseudobulge structure, as does the
positive $B_6$ term.  
Given the unrelaxed nature of the system, this bulge may still be growing. 
At small radii ($R_{\rm eq} < 3\arcsec$), 
knots along the major-axis give a disturbed appearance to the isophotes. 
Some of this has been captured by the model, 
evident by the spikes in the fourth- and sixth-order Fourier harmonic 
profiles (Figure~\ref{Fig-Prof}). 
while most has been left behind in the residual image. 
The light profile in Figure~\ref{Fig-Prof} also displays a prominent upturn within the inner
arcsecond.  This may not originate from a single nuclear star cluster, 
but seems to reflect the unrelaxed state of play at the center of the galaxy
where a few knots are seen.

The somewhat stretched star
cluster in NGC~4424, which we shall refer to as `Nikhuli' (see section~\ref{Sec-BH-Nik}), 
is drawn out in the direction of the galaxy's center.  
From the WFC3/IR F160W residual image (galaxy minus model), we measure this elongated
source to have a flux of $879\pm30$ electrons per second.  Using a WFC3/IR
F160W zero-point\footnote{The sensitivity of the WFC3/IR detector has been
  remarkably stable over the past decade \citep{2020wfc..rept....5K} and
  offers a 1$\sigma$ photometric accuracy/repeatability of 1.5\% (0.015
  mag) \citep{2019wfc..rept....7B}.} of 25.946 mag \citep[AB photometric
  system:][]{2013ApJS..209....3K}, this corresponds to an apparent magnitude
of 18.59$\pm$0.04 mag (AB).  Using a distance modulus of 31.08$\pm$0.29 gives
an absolute magnitude of $-$12.49$\pm$0.29~mag.  With an F160W absolute
magnitude for the Sun of $+$4.60~mag \citep[AB mag:][]{2018ApJS..236...47W},
the stellar luminosity is $(6.88\pm1.85)\times10^6\,$L$_{\odot,F160W}$.
Assuming an F160W stellar mass-to-light ratio of 0.5 \citep[][see their
  Figure~A1]{2009MNRAS.397.2148G}, the stellar mass is 
$(3.44\pm0.93)\times10^6\,M_{\odot}$.  The Galactic extinction at 1.6 $\mu$m, in the direction of
NGC~4424, is just 0.01 mag \citep{2011ApJ...737..103S}.  From the 3-band
composite color image in \citet{2016ApJ...819...31G}, it also appears that
dust intrinsic to NGC~4424 is not a significant issue at the location of Nikhuli.

\subsubsection{Color}

Given that Nikhuli is also visible in an F814W image, it rules out the possibility that the
detection in the F160W image was solely due to Fe\,{\footnotesize II} 1.64 $\mu$m
line emission from an ionized bubble around NGC~4424 X-3.  The elongated
morphology also disfavours such an interpretation.  We used Gaussian smoothing
to degrade the spatial resolution of the WFC3 UVIS channel F814W 
image---obtained from  
the same {\it HST} Proposal (ID 12880. PI: A.Riess) as the F160W image 
--- to match the seeing in the F160W image.  The (galaxy plus star
cluster) light in a 3$\farcs$4 (272~pc) by 1$\farcs$5 (120~pc) ellipse
centered on the star cluster has a mean F814W$-$F160W color equal to
1.60$\pm$0.10 mag.  This color is slightly redder than the surrounding mean
galaxy color of 1.44$\pm$0.04 mag, and may be indicative of an older stellar
population.  This would not be unusual.  For example, the star cluster in the
Milky Way dwarf spheroidal satellite Eridanus II is also thought to be old
\citep{2021MNRAS.505.2074A}.  

Because the (color of the) galaxy light can dominate over the (color of the)
star cluster light, the F814W$-$F160W galaxy (plus star cluster) color image is
not optimal for showing the color of Nikhuli. We have, therefore, attempted to provide 
a color residual image rather than a color galaxy 
image.  To do this, we subtracted the F160W residual
image from the F814W residual image.  By design, most of the pixels will have
intensities close to zero in these residual images.  
As such, the average `color' in the color residual
image will be close to zero rather than equal to the galaxy color. 
Furthermore, 
roughly half of the pixels in the F814W and F160W residual images have a 
negative intensity, especially where there was obscuration from dust.  
This can be problematic because the traditional color, given by the
ratio of intensities, $-2.5\log(I_1/I_2)$, is ill-defined when $I_1/I_2$ is
negative, and the color image will be noisy where either $I_1$ or $I_2$ are close
to zero. To mitigate against this, in the residual images  
we added a pixel intensity of 
1 to those pixels with a positive intensity, 
and we subtracted a pixel intensity of
1 to those pixels with a negative intensity.
For each of these modified residual images, 
we then calculated the pseudo-magnitudes $-2.5\log(I+1)$ for those pixels with
positive values of $I$, and $+2.5\log(|I-1|)$ for those pixels with negative values of $I$.
We then used mag$_1 -$mag$_2$ as our proxy for each pixel's color. 
Due to the processes of image rotation, alignment, pixel-size matching, 
seeing matching, plus 
our addition of the `bias level' to the intensity, our pseudo-color image for the residual
structure is not perfect, but it is still informative.

\begin{figure*}
$
\begin{array}{cc}
 \includegraphics[angle=0, trim=0cm 0.0cm 0cm 0cm, width=1.0\columnwidth]{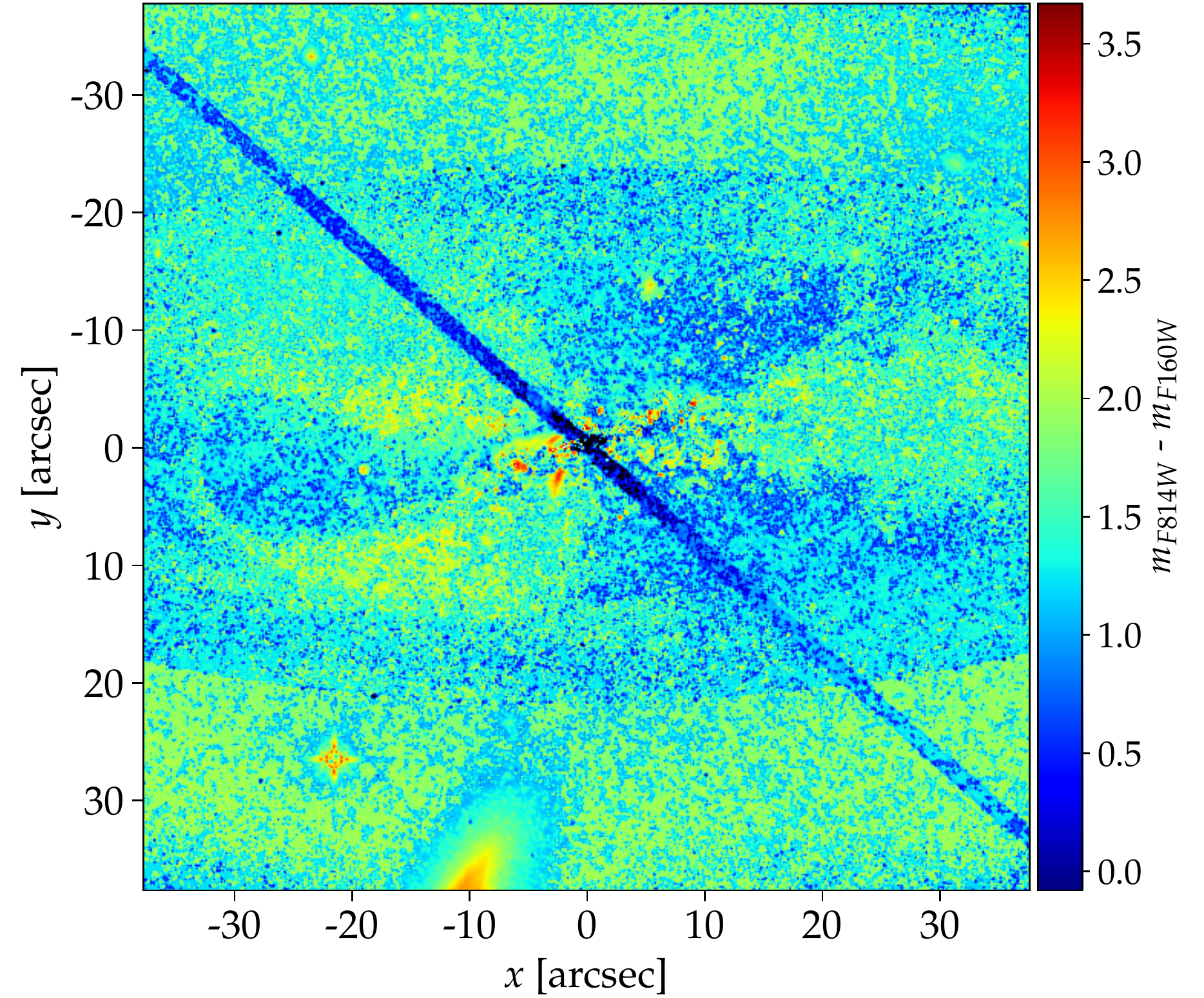} &
 \includegraphics[angle=0, trim=0cm 0.0cm 0cm 0cm, width=1.0\columnwidth]{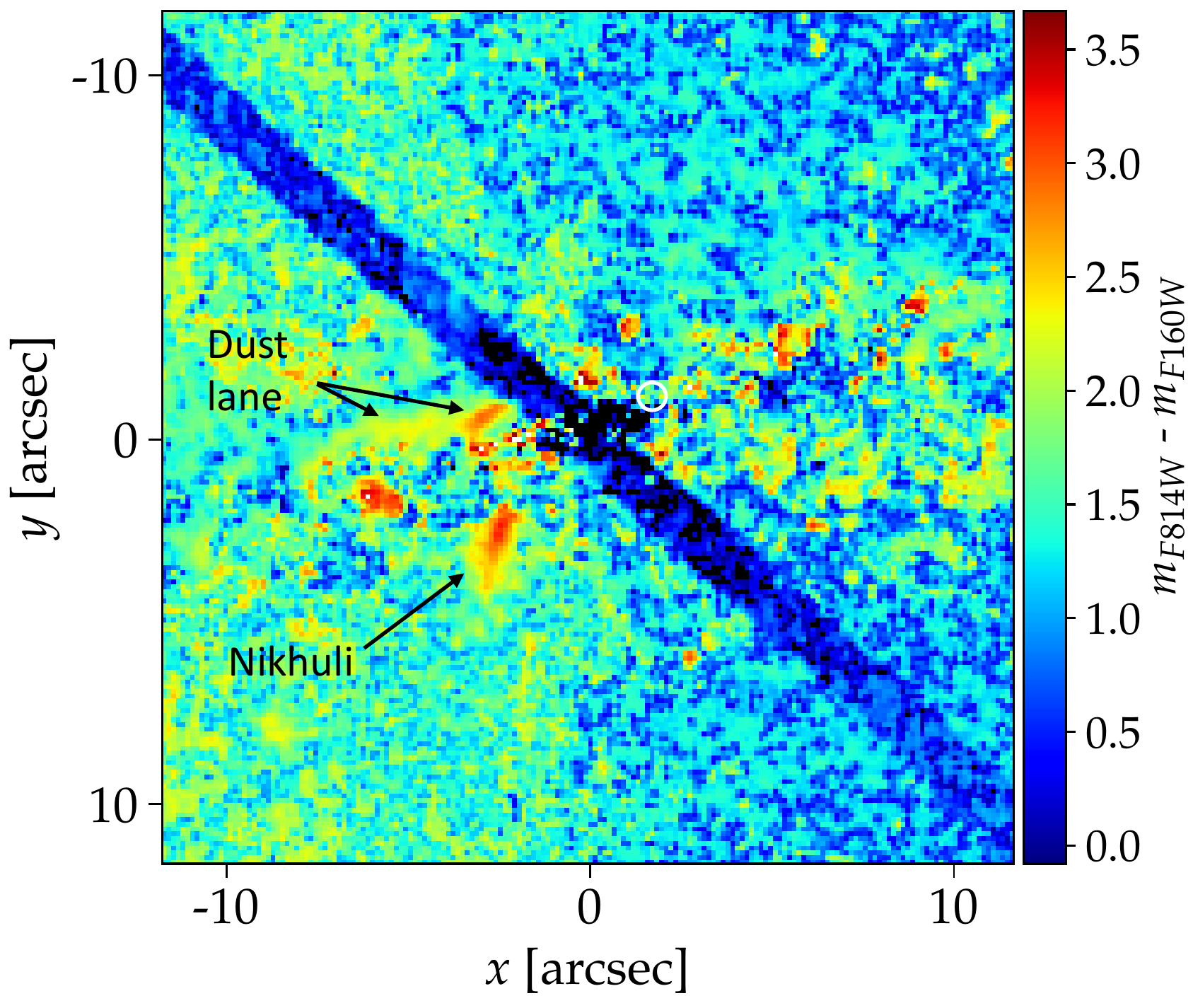} \\
\end{array}
$ 
\caption{Pseudo-color residual image, and central zoom (right), 
  after removing much of the mirror-symmetric component of 
  NGC~4424 about its major axis.  The WFC3-IR/F160W residual image
  (Figure~\ref{Fig_4424_resid}) has been subtracted from the WFC3-UVIS/F814W residual
  image.  The WFC3-UVIS chip gap is evident by the diagonal blue
  stripe.  The galaxy center has been circled in white. 
  Residual flux shows up as red or blue
  depending on whether the relative intensity in the F160W to F814W residual
  image is high or low.  See section~\ref{Sec_Opt} for details.}
\label{Fig_color}
\end{figure*}

The prominent stripe seen in the  pseudo-color residual image (Figure~\ref{Fig_color})
is from the 1$\farcs$24-wide WFC3 UVIS detector chip gap in the F814W image.  
The yellow, three-pronged, fan pattern is due to Isofit more successfully
capturing and removing  this real galaxy structure from the F160W image than
from the F814W image. 
While much of the blue color within the fitted quasi-elliptical 
region is noise, arising from zero minus zero, the image does reveal the
relative intensity of F160W to F814W light in several features across the field. 
One can see that the light from Nikhuli (associated with emission in
Figure~\ref{Fig_4424_resid}) has a relatively high F160W to F814W intensity
ratio.  One can also see that the dust lane, 
associated with obscuration in Figure~\ref{Fig_4424_resid}, and a studied feature
in some spiral galaxy bars \citep[e.g.,][]{1992MNRAS.259..345A}, 
also appears red. 
In contrast, an inner 3$\farcs$2 ($\sim$260 pc) region of the disk,
immediately east of the galaxy center, appears to have a relatively low F160W
to F814W intensity ratio. This is indicative of recent/ongoing star formation.
Unfortunately, the colors seen in Figure~\ref{Fig_color} are not perfect. 
A hindrance to a more quantitatively useful color residual image is the
apparent lack of structure in the F814W image relative to the F160W image,
which tends to amplify the signal from features in the F160W image. 
In addition, a galaxy model was independently fit to both the F160W and F814W
images, and thus sightly different models have been subtracted from each
image, affecting the build of the color image.

\begin{figure}
\includegraphics[angle=00,width=1.0\columnwidth]{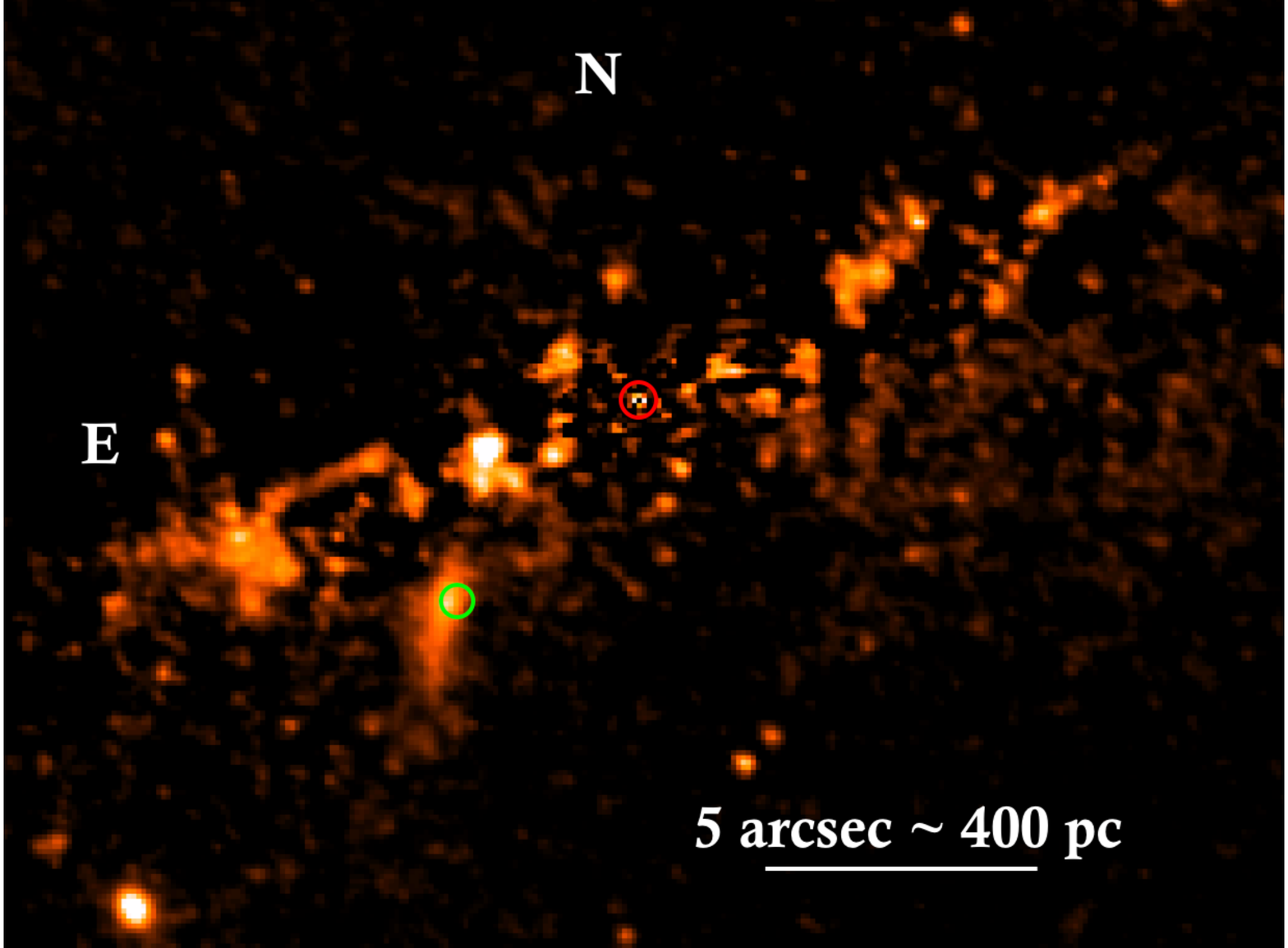}
\caption{Residual image from Figure~\ref{Fig_4424_resid}, with a slightly
  different contrast and with the center of the X-ray point-source NGC~4424
  X-3 shown by the green circle.  It has a radius of $0.\arcsec3$ and 
  encapsulates the 90\% uncertainty on the X-ray source position after mapping
  into the {\it HST} image. The red circle marks the center of the galaxy NGC~4424.
}
\label{Fig_centre}
\end{figure}

\subsection{X-Ray data}

\subsubsection{Chandra X-Ray Observatory Data}
\label{Sec_Chandra}

Using the 14.87~ks {\it CXO} Advanced CCD Imaging Spectrometer (ACIS)
exposure (ObsID: 19408. PI: Soria) taken on 2017-04-17, \citet[][see their
  sections 4.2]{2018A&A...620A.164B} reported on the obscured X-ray source
NGC\,4424 X-3, located 4$\farcs$9  ($\approx$390~pc) to the
southeast of the nucleus.  The {\it CXO} enables accurate subarcsecond
localizations of faint point-like sources and NGC\,4424 X-3 is coincident with
the star cluster.\footnote{The {\it CXO/HST} alignment was refined with the help of the 
point-like {\it CXO} source associated with the nucleus of the background galaxy at
R.A.=12:27:12.16, decl.=+9:24:35.55 and with a quasar at R.A.=12:27:12.85, 
decl.=+9:24:13.13.} 
Figure~\ref{Fig_centre} shows the X-ray point-source at 
R.A.=12:27:11.801, decl.=+9:25:10.71, with a red circle denoting the 
90\% radius = 0$\farcs$3.  

We have reanalyzed the {\it CXO} data following the detailed prescription
provided in \citet{2021Graham} and \citet{Soria2021}. Briefly, we used the
Chandra Interactive Analysis of Observations ({\sc ciao}) Version 4.12
software package \citep{2006SPIE.6270E..1VF} with the Calibration Database
Version 4.9.1, to reprocess the event files ({\sc ciao} task {\it
  chandra\_repro}), and the display package {\sc ds9}
\citep{2003ASPC..295..489J} 
for image processing, centroiding, and selection of source and background
regions. NGC\,4424 X-3 is located close to the ACIS aimpoint, which makes the
point spread function sharper and reduces the positional uncertainty of the
centroid even with a small number of counts. To take advantage of this, we
also extracted images at sub-pixel resolution (0.5 pixels, corresponding to
$\approx$0$\farcs$25), with {\it dmcopy}, and used the 0.3--7 keV
sub-pixel image for centroiding.

We used a source extraction radius of 1$\farcs$5 (3 pixels) around
the source centroid; the expected encircled energy within this radius is
$\approx$95\%. For the background, we used the annulus between
2$\farcs$5 and 10$\arcsec$. There are 8 X-ray counts within
the source region (in fact, all within 1$\farcs$0 of the centroid
position), in the 0.3--7 keV band: none are in the soft band (0.3--1 keV), 1
is in the medium band (1--2 keV), and 7 are in the hard band (2--7
keV). Splitting the energy band in a different way, we see that all 8 counts
are $>$1.5 keV. Although the number of counts is small, the detection is
highly significant, because the background is very low; based on the local
average, we expect only $\approx$0.1 background counts in the soft band, in
the source region; $\approx$0.2 background counts in the medium band; and
$\approx$0.1 background counts in the hard band. For $N=8$ detected counts and
$\approx$0.4 expected background counts in an integration time of 14.87 ks, 
the 90\% Bayesian confidence interval 
is $\approx$4--13 net counts, and the 99\% interval is $\approx$2--17 net
counts \citep{1991ApJ...374..344K}. 
The net 0.3--7 keV count rate (including the PSF
correction to an infinite radius) is $0.56^{+0.42}_{-0.29}\times 10^{-3}$ ct
s$^{-1}$ (90\% Bayesian confidence interval).

\begin{figure*}
$
\begin{array}{cc}
 \includegraphics[angle=0, trim=0cm 0.0cm 0cm 0cm, width=1.0\columnwidth]{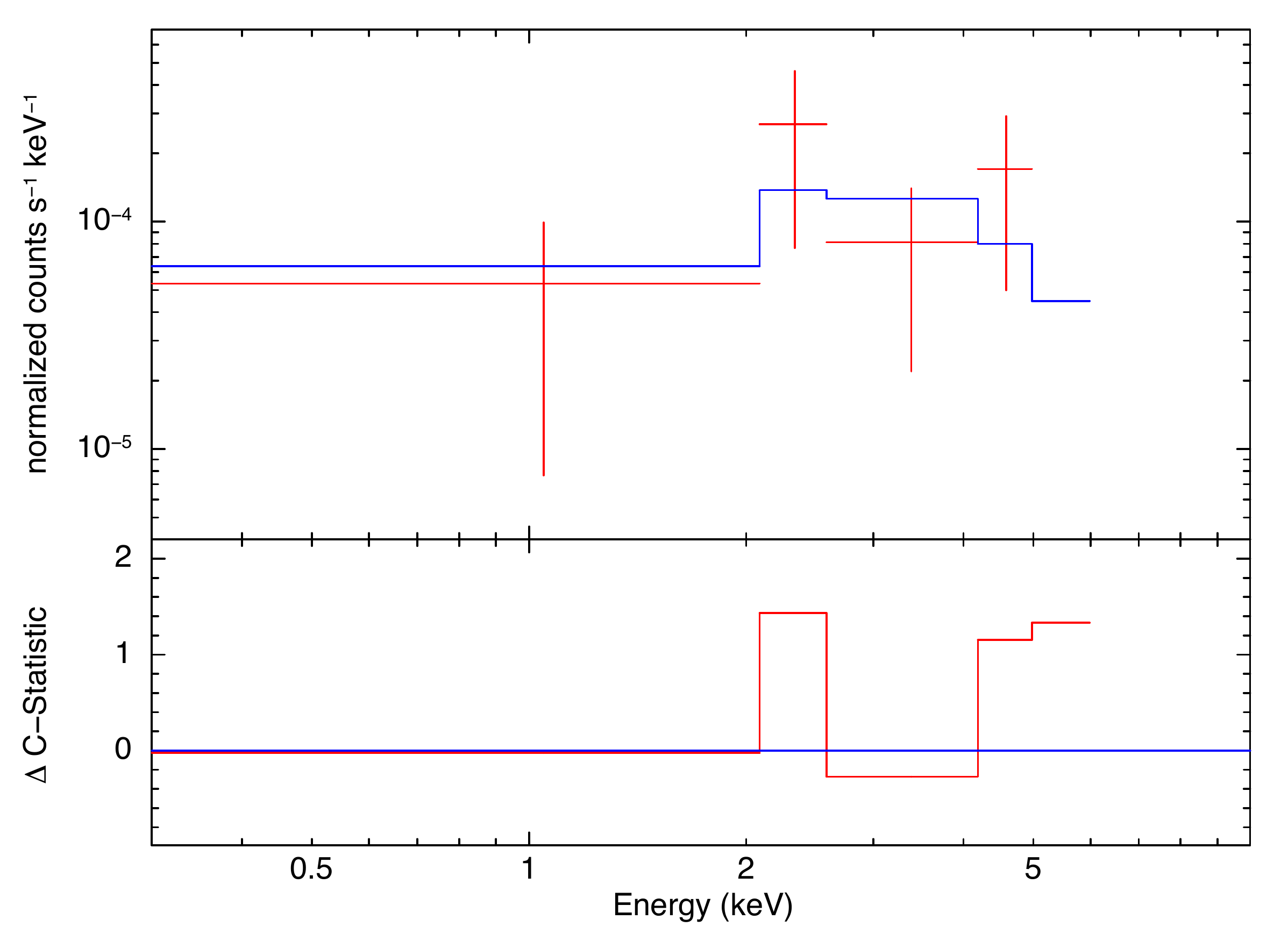} &
 \includegraphics[angle=0, trim=0cm 0.0cm 0cm 0cm, width=0.97\columnwidth]{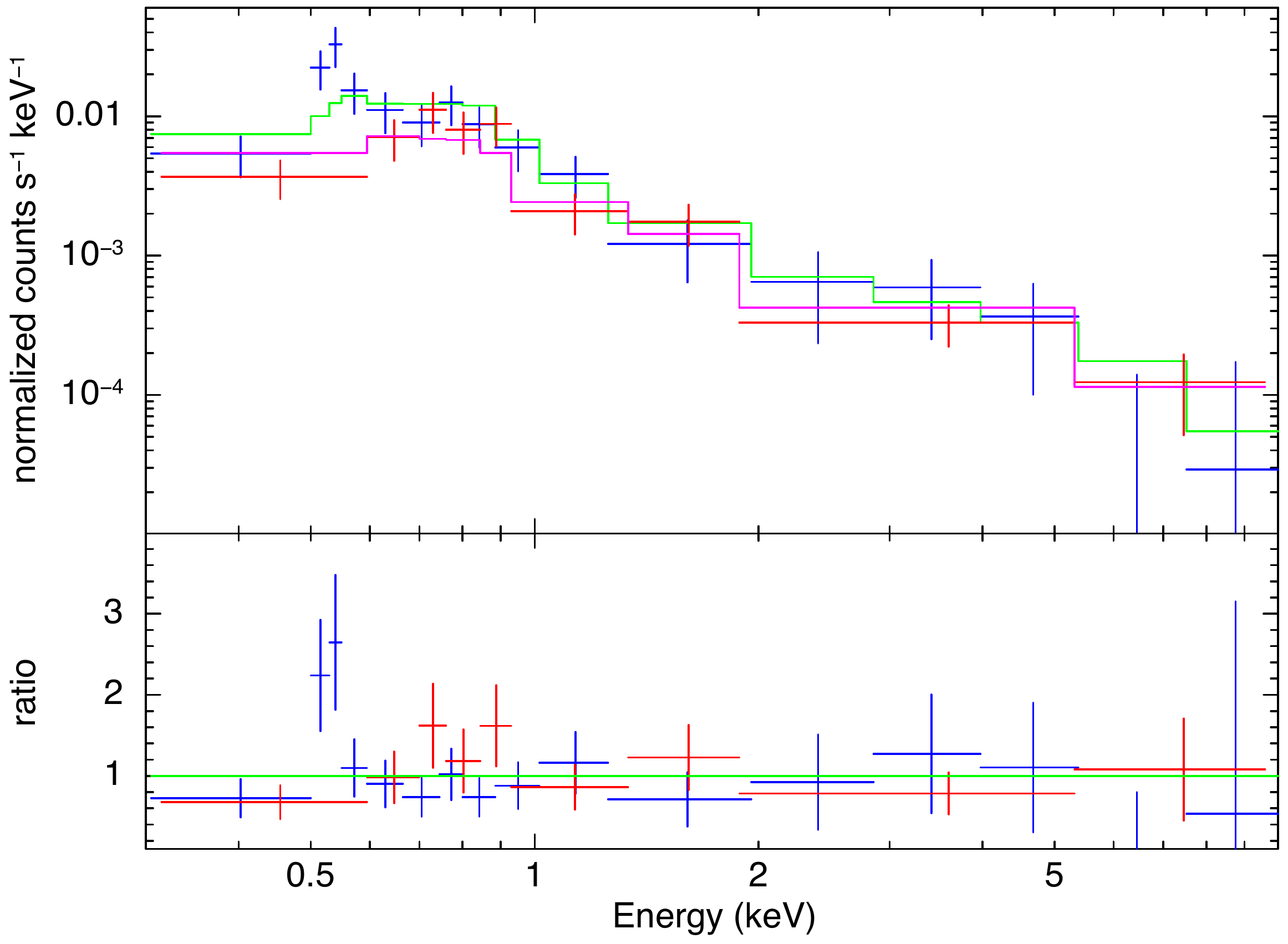} \\
\end{array}
$ 
\caption{Left panel: CXO/ACIS-S spectrum of NGC~4424 X-3 in Nikhuli, 
with a $\Gamma=1.7$ power-law model (blue line). The data points have 
been grouped to a signal-to-noise $>$1.8 for plotting purposes only.
The fit was done on the individual counts, using Cash statistics. 
Right panel: 
Combined XMM-Newton/EPIC spectra of NGC\,4424 X-3 in 2010 and 2017, along
with the best-fitting models and the data/model ratios. Red data points are for
the 2010 spectrum; the corresponding best-fitting model is plotted in
magenta. Blue data points are for the 2017 spectrum; the corresponding
best-fitting model is plotted in green. The EPIC spectra were grouped to 1
point per bin and fitted with the Cash statistics; the results were later
rebinned to a signal-to-noise ratio $>$2.5 and a maximum number of grouped
channels = 50, for plotting purposes only. The photon index (locked between
the two epochs) of the power-law component is $\Gamma = 1.7^{+0.9}_{-0.7}$;
the de-absorbed 0.3--10 keV luminosity of the power-law (i.e., not
including the thermal plasma emission) is $\approx$10$^{39}$ erg s$^{-1}$ in
the 2010 EPIC spectrum and $\approx$2.2 $\times 10^{39}$ erg s$^{-1}$ in the
2017 spectrum. 
}
\label{Fig-X_spec}
\end{figure*}

We tried to infer more information on the flux and luminosity of X-3 in two
different ways: 
from the energy of the detected photons (using the {\sc CIAO} task {\it srcflux}) and by
fitting a spectral model (using the {\sc xspec} \citep{1996ASPC..101...17A}
package version 12.11.0). 
First, we applied the {\it srcflux} task to the reprocessed
event file to obtain a model-independent estimate of the observed, 
(H\,{\footnotesize I} gas)-dimmed, flux\footnote{{\it srcflux}'s 
model-independent flux is based on the directly measured rate and energies of
the detected photons, taking into account the quantum efficiency and effective
area within the source region. Instead, the model-dependent flux is the flux
of the best-fitting model to the observed photon distribution. They are
alternative approximations to the `true' observable (absorbed) flux. We used
{\it srcflux} to calculate the model-independent flux, and {\sc xspec} to determine the
model-dependent flux. From {\sc xspec}'s model-dependent flux we then computed the
unabsorbed flux, and hence estimated the emitted luminosity.} in the 
`broad' ACIS band (0.5--7 keV). Second, despite the small number of counts,
we extracted a spectrum and its associated response and ancillary response
files with {\it specextract} (Figure~\ref{Fig-X_spec}). 
We regrouped the spectrum to 1 count per bin,
and fitted it with simple models (power-law and disk-blackbody) using 
{\sc xspec}.  
We used the \citep{1979ApJ...228..939C} statistics for the fit ({\it cstat} in {\sc xspec}).

The main purpose of our spectral analysis was obviously not to determine a
complex fitting model, but at least to estimate what intrinsic column density
and de-absorbed luminosity are consistent with the peculiar energy
distribution of the few observed photons (all above 1.5 keV). Assuming a
power-law model (the simplest option in the absence of further information)
with a photon index $\Gamma = 1.7$ \citep[a standard, reference choice for the hard
emission of accreting compact objects, e.g.][]{2009MNRAS.399.1293M, 2011A&A...530A..42C, 2011A&A...530A.149Y, 2013ApJ...773...59P, 2015MNRAS.447.1692Y}, we estimate that an intrinsic
absorbing column $N_{\rm {H,int}} \gtrsim 10^{22}$ cm$^{-2}$ (much higher than
the Galactic line-of-sight column density $N_{\rm {H,Gal}} = 1.5 \times
10^{20}$ cm$^{-2}$; \citep{2016A&A...594A.116H}) is needed to explain the
energy distribution of the detected photons (Table~\ref{Tab_Sum}); a similarly
high absorption is required if the spectral model is a disk-blackbody with an
assumed temperature of 1.0 keV. This rough estimate of the intrinsic
absorption permits a more realistic conversion of count rate and flux to
intrinsic luminosity. For the assumed\footnote{The
power-law slope, aka photon index, $\Gamma$, is around 1.7 for Eddington
ratios $\approx10^{-5}$--$10^{-2}$ \citep[e.g.][]{2009MNRAS.399.1293M, 
2011A&A...530A..42C, 2011A&A...530A.149Y, 2013ApJ...773...59P, 2015MNRAS.447.1692Y}.
For Eddington ratios less than $10^{-5}$, $\Gamma$ steepens to around 2.1, and for Eddington ratios
  greater than $10^{-2}$, $\Gamma$ steepens to around 2.5.} $\Gamma = 1.7$ 
power-law model\footnote{With $\Gamma = 1.7$, one has that
$L_{0.5-8\,{\rm keV}} = 1.075\,L_{0.5-7\,{\rm keV}}$, and
$L_{0.5-10\,{\rm keV}} = 1.207\,L_{0.5-7\,{\rm keV}}$.
Similarly,
$L_{2-10\,{\rm keV}} = 0.725\,L_{0.5-8\,{\rm keV}}$, and
$L_{0.5-10\,{\rm keV}} = 1.123\,L_{0.5-8\,{\rm keV}}$.}, the
unabsorbed 0.3--10 keV luminosity is $L_{0.3-10} \approx 10^{+14}_{-6}
\times10^{38}$ erg s$^{-1}$.  For the disk-blackbody model, $L_{0.5-8} \approx
9^{+14}_{-5} \times10^{38}$ erg s$^{-1}$ (Table~\ref{Tab_Sum}). As a further
check of these results, we input the net count rates in the hard band (derived
earlier with {\it srcflux}) into the {\sc ciao} version of the Portable,
Interactive Multi-Mission Simulator ({\sc pimms}) version
4.11a\footnote{\url{https://cxc.harvard.edu/toolkit/pimms.jsp}}, with response
functions for Chandra Cycle 18. We recovered the same estimates for the
intrinsic flux (corresponding to an extrapolated 0.3--10 keV luminosity of
about $10^{39}$ erg s$^{-1}$ for plausible spectral slopes), and the same high
values of $N_{\rm {H,int}} \gtrsim 10^{22}$ cm$^{-2}$ to remove all photons
below $\approx$1.5 keV.
The {\it CXO} image is shown in Figure~\ref{Fig-NGVS}.

We do not have enough counts and bandwidth coverage to decide between a
power-law model and a curved (e.g., disk-blackbody) model, even with the Cash
statistics. A power-law model is the expected spectral shape for an IMBH,
which would be in the low/hard state at a luminosity of $\approx$10$^{39}$ erg
s$^{-1}$ ($\sim$10$^{-3}$ times the Eddington luminosity for a
$10^{4}\,M_{\odot}$ 
black hole). However, the luminosity of $\approx$10$^{39}$ erg s$^{-1}$ is
also consistent with a stellar-mass black hole at the top of the high/soft
state, with a typical temperature of $\approx$1 keV and an inner disk radius
$\sim$50--100 km.

\begin{table}
\caption{Best-fitting parameters of the Chandra/ACIS spectrum of
  NGC~4424 X-3.}
\vspace{-0.3cm}
\label{Tab_Sum}
\begin{center}  
\begin{tabular}{lc}
 \hline
\hline \\[-8pt]
  Model Parameters      &      Values \\
\hline\\[-9pt]
\multicolumn{2}{c}{Model-independent 0.5--7 keV flux from {\it srcflux}}\\
\hline\\[-9pt]
Net Rate ($10^{-3}$ ct s$^{-1}$)  & $0.56^{+0.42}_{-0.29}$ \\[4pt]
$f_{0.5-7}$ ($10^{-14}$ erg cm$^{-2}$ s$^{-1}$)
       & $ 0.79^{+0.59}_{-0.40}$\\[4pt]
\hline\\[-9pt]
\multicolumn{2}{c}{{\it phabs} $\times$ {\it phabs} $\times$ {\it power-law}}\\
\hline\\[-9pt]
   $N_{\rm {H,Gal}}$   ($10^{22}$ cm$^{-2}$)    &    [0.015]        \\[4pt]
   $N_{\rm {H,int}}$   ($10^{22}$ cm$^{-2}$)   &  $ 2.7^{+3.8}_{-1.9}$
\\[4pt]
   $\Gamma$      &  [1.7]     \\[4pt]
   $N_{\rm {po}}$ ($10^{-6}$ photons keV$^{-1}$ cm$^{-2}$ s$^{-1}$ at 1 keV)
              &  $ 4.5^{+6.1}_{-2.8}$  \\[4pt]
   C-stat/dof     &      $2.9/7$           \\[4pt]
   $f_{0.5-7}$ ($10^{-14}$ erg cm$^{-2}$ s$^{-1}$)$^a$
       & $ 1.1^{+1.0}_{-0.6}$\\[4pt]
   $L_{0.3-10}$ ($10^{38}$ erg  s$^{-1}$)$^b$
      & $10.0^{+13.7}_{-6.1}$\\[2pt]
   \hline\\[-9pt]
\multicolumn{2}{c}{{\it phabs} $\times$ {\it phabs} $\times$ {\it diskbb}}\\
\hline\\[-9pt]
   $N_{\rm {H,Gal}}$   ($10^{22}$ cm$^{-2}$)    &    [0.015]        \\[4pt]
$N_{\rm {H,int}}$   ($10^{22}$ cm$^{-2}$)   & $3.4^{+4.6}_{-2.3}$  \\[4pt]
   $kT_{\rm in}$ (keV)     &  [1.0] \\[4pt]
   $N_{\rm {dbb}} $  ($10^{-3}$ km$^2$)$^c$ &  $ 1.3^{+2.1}_{-0.8}$\\[4pt]
   $R_{\rm {in}}\sqrt{\cos \theta} $  (km)$^d$ & $70^{+43}_{-28}$\\[4pt]
   C-stat/dof     &     $3.4/9$ \\[4pt]
   $f_{0.5-7}$ ($10^{-14}$ erg cm$^{-2}$ s$^{-1}$)$^a$
       & $ 0.81^{+0.69}_{-0.42}$\\[4pt]
   $L_{0.3-10}$ ($10^{38}$ erg s$^{-1}$)$^b$
      & $ 8.5^{+13.5}_{-5.4}$\\[2pt]
\hline
\vspace{-0.5cm}
\end{tabular}
\end{center}
\begin{flushleft}
{\bf Notes.} Error ranges are 90\% confidence limits for 1 interesting
  parameter. Values in brackets were frozen in the fit. \\
$^a$ absorbed model flux in the 0.5--7 keV band.\\
$^b$ isotropic de-absorbed luminosity in the 0.3--10 keV band, defined as
  $4\pi d^2$ times the de-absorbed 0.3--10 keV model flux.\\
$^c$ $N_{\rm {dbb}} = (r_{\rm{in}}/d_{10})^2 \cos \theta$, where
  $r_{\rm{in}}$ is the apparent inner disk radius in km, $d_{10}$ the distance
  to the source in units of 10 kpc (here, $d_{10} = 1640$), and $\theta$ is
  our viewing angle ($\theta = 0$ is face-on).\\
$^d$ $R_{\rm {in}} \approx 1.19 r_{\rm in}$ for a standard disk
  \citep{1998PASJ...50..667K}.
\end{flushleft}
\end{table}

\subsubsection{XMM-Newton data}

The field around NGC\,4424 X-3 was also observed with the {\it XMM-Newton}
European Photon Imaging Camera - Metal Oxide Semi-conductor (EPIC-MOS) on two
occasions: on 2010 June 13 (Obs.ID: 0651790101; Revolution: 1925) and 2017
December 5 (Obs.ID: 0802580201; Revolution: 3295), see \citet[][their
  Figure~5]{2018A&A...620A.164B}.  We downloaded the data from the HEASARC
archive and reprocessed them with the Science Analysis Software ({\sc sas})
\citep[version 17.0.0:][]{2004ASPC..314..759G}.
We rebuilt event files for the pn and MOS cameras with {\it
  epproc} and {\it emproc}, respectively. We filtered the event files to
remove intervals of high or rapidly flaring particle background; we selected
only intervals with a background RATE parameter $<$0.35 for MOS1 and MOS2, and
$<$0.5 for the pn, at channels 10000 $<$ PI $<$ 12000. After this filtering
step, the good-time-interval for Obs.ID 0651790101 in the source extraction
region was 13.8 ks for the pn, and 20.6 ks for each of the two MOS cameras;
for Obs.ID 0802580201, the live time was 11.1 ks for the pn and 18.9 ks for
the MOSs. We then filtered the event files with the standard conditions
``(FLAG$==$0) \&\& (PATTERN$<=$4)'' for the pn, and ``(\#XMMEA\_EM \&\&
(PATTERN$<=$12)'' for the MOSs (corresponding to single and double events).

We defined a circular source region of radius 15$\arcsec$: this is slightly
smaller than usual for EPIC point-like sources, but we wanted to reduce
contamination of the faint point-like source from the surrounding diffuse
emission expected from the hot gas in the central region of this star-forming
galaxy. (This is obviously not an issue in Chandra, given its much higher
spatial resolution). We defined local background regions seven times larger
than the source region. For each of the two observations, we used {\it
  xmmselect} to extract individual spectra for the pn and the MOSs; we built
the associated response and ancillary response files with {\it rmfgen} and
{\it arfgen}. Finally, we combined the pn and MOS spectra of each observation
with {\it epicspeccombine}. We grouped the two resulting EPIC spectra to a
minimum of 1 count per bin, and fitted them with {\sc xspec}
\citep{1996ASPC..101...17A} version 12.11.0, using the Cash statistics
\citep{1979ApJ...228..939C}. We chose the Cash statistics because the combined
spectra do not have enough counts for a meaningful $\chi^2$ fitting: we
estimate $\approx$190 net counts in the EPIC spectrum from Obs.ID 0651790101,
and $\approx$120 net counts from Obs.ID 0802580201.

Both spectra are well fitted with thermal plasma emission, dominating below
$\approx$1 keV, and a power-law component, dominating at higher energies. In
particular, the soft component is consistent with multi-temperature thermal
plasma. When fitted with a two {\it apec} components, the two temperatures are
$\approx$0.2 keV and $\approx$0.7 keV. This is the characteristic temperature
distribution of diffuse hot gas in star-forming galaxies
\citep{2009MNRAS.394.1741O, 
2012MNRAS.426.1870M}. Thus, we interpret the power-law component as the
emission from the accreting compact object, and the soft component as a local
enhancement of diffuse emission, tracer of recent star formation (perhaps
associated with the satellite merger). The soft thermal component must be seen
through a much lower column density than estimated from the hard colors of
the Chandra detection.

Based on this tentative interpretation, and in order to limit the number of
free parameters, we froze the temperatures of the two {\it apec} components at
0.2 keV and 0.7 keV, and added an additional intrinsic absorbing column
density of $10^{22}$ cm$^{-2}$, seen only by the power-law component. We also
kept the {\it mekal} normalizations and the power-law photon index locked
between the two epochs. We obtain a good fit (C-statistics of 496.8/497) for a
photon index $\Gamma = 1.7^{+0.9}_{-0.7}$. The 0.3--10 keV deabsorbed
luminosity of the power-law component (which corresponds to the point-like
Chandra source, in our model interpretation) roughly doubled from
$\approx$10$^{39}$ erg s$^{-1}$ in 2010 to $\approx$2.2$\times 10^{39}$ erg
s$^{-1}$ in 2017 (0.5--8 keV luminosities of  $\approx$8$\times 10^{38}$ erg
s$^{-1}$ and $\approx$1.8$\times 10^{39}$ erg s$^{-1}$, respectively). 

In passing we note that because 
the X-ray source has not faded over 7 years, 
it is unlikely to have originated from a 
stellar tidal disruption event (TDE) from one of the stars in the star
cluster approaching too close to the putative black hole \citep{2006MNRAS.372..467B,
  2015JHEAp...7..148K, 2018NatAs...2..656L}.

\section{Black hole mass estimates}\label{Sec_BH}

NGC~4424 might harbour two massive black holes,
possibly with one yet to settle to the center of this 
merged system \citep{2005LRR.....8....8M, 2012MNRAS.423.2533B,
  2018ApJ...857L..22T}, thereby explaining NGC~4424 X-3 in the
elongated, off-center star cluster.  In this scenario, the second (speculated) 
massive black hole is quiescent, residing at the center of NGC~4424.
 
In what follows, we first report on the predicted mass for such a galaxy-centric black
hole.  We then provide a prediction for the mass of the black hole in the star
cluster Nikhuli.

\subsection{NGC 4424}
 
Using a revised distance modulus of 31.080$\pm$0.292 ($=16.4\pm0.8$~Mpc),
rather than 30.6$\pm$1.0, NGC~4424 has a revised (total) stellar mass $\log
M_{\rm *,gal}=10.0\pm0.1$ \citep[][see their Appendix]{2019MNRAS.484..814G}.
This mass is based on the near-IR $K^{\prime}$-band (2.2 $\mu$m) galaxy
magnitude of 8.86 mag from GOLD
Mine\footnote{\url{http://goldmine.mib.infn.it/}} \citep{2003A&A...400..451G},
plus our assumed 0.10 mag uncertainty and a $K^{\prime}$-band stellar
mass-to-light ratio for the galaxy of $M/L_{K^{\prime}}=0.62\pm0.09$.  This
galaxy stellar mass implies a central black hole mass of $\log M_{\rm bh} =
4.8\pm0.8$ (based upon a symmetrical Bayesian analysis of the $M_{\rm
  bh}$--$M_{\rm *,gal}$ relation for late-type galaxies) or $5.1\pm0.8$ (based
upon a symmetrical bisector regression for late-type galaxies using the
modified FITEXY routine).  See \citet{2018ApJ...869..113D} and
\citet{2019MNRAS.484..814G} for details.

The merger event in NGC~4424 may have 
elevated the central stellar velocity dispersion, reported to be $57.0\pm8.6$
km s$^{-1}$ 
by \citet{2009ApJS..183....1H}.  Although, this is perhaps in line with the
(merger-induced increase to the) stellar mass as it results in a consistent prediction for
the black hole mass of $\log M_{\rm bh} = 5.0\pm0.8$
\citep{2017MNRAS.471.2187D}. 
The past/ongoing merger has, however, distorted the spiral
pattern \citep{1996AJ....111..152K,2006AJ....131..747C}, likely
accounting for the discrepant prediction, see Table~\ref{Tab_IMBH}, 
for the black hole mass obtained using the 
spiral arm pitch angle \citep[16.9$\pm$2.4 degrees:][]{2019MNRAS.484..814G}
and the $M_{\rm bh}$--$\phi$ relation in \citet{2017MNRAS.471.2187D}.

\begin{table}
\centering
\caption{Black hole mass predictions for the center of NGC~4424.}\label{Tab_IMBH}
\begin{tabular}{cccc}
\hline
$\log M_{\rm bh}$ ($M_{\rm *,total}$) & $\log M_{\rm bh}$ ($\phi$) & $\log M_{\rm bh}$ ($\sigma$)  & $\overline{\log M_{\rm bh}}$ \\
\hline
4.8$\pm$0.8                    &   6.7$\pm$0.5       &  5.0$\pm$0.8                           &    4.9$\pm$0.6         \\
\hline
\end{tabular}

\begin{flushleft}
The predicted, central black hole masses shown here are in units of solar
mass.  As explained in Section~\ref{Sec_BH}, 
the predictions were derived from the galaxy's total stellar mass, spiral arm pitch angle
and stellar velocity dispersion. 
The pitch angle, however, is considered unreliable 
for predicting the black hole mass in this post-merger galaxy, 
and it is therefore not used to calculate the error-weighted
mean black hole mass, $\overline{\log M_{\rm bh}}$, shown in the final column.
These masses should not be confused with the predicted, logarithmic black hole mass in Nikhuli,
equal to $4.8\pm1.6$ (see the end of section~\ref{Sec_BH}). 
\end{flushleft} 
\end{table}

The error-weighted mean of the logarithm of the estimated black
hole masses can be calculated via 
\begin{equation}
\overline{\log M_{\rm bh}} = \frac{ \sum_{i=1}^N w_i \log M_{{\rm bh},i}
}{\sum_{i=1}^N w_i}, 
\end{equation}
with inverse-variance weighting\footnote{This gives
  the `maximum likelihood estimate' for the mean of the probability
  distributions under the assumption that they are independent and normally
  distributed with the same mean.}  such that $w_i = 1/(\delta \log M_{{\rm
    bh},i})^2$.  The associated 1$\sigma$ standard error is given by 
\begin{equation}
\delta \, \overline{\log M_{\rm bh}} = \sqrt{ 1 / \sum_{i=1}^N w_i  }.
\end{equation}   

Based on the stellar mass and velocity dispersion of NGC~4424, 
it is predicted to possess a central 
intermediate mass black hole with $\overline{\log M_{\rm bh}}=4.9\pm0.6$ dex. 
The benefit of such an error-weighted mean is that the uncertainty is less
than the near-order of magnitude 
uncertainty on the individual black hole mass estimates. 

\subsection{Nikhuli}
\label{Sec-BH-Nik}

The elongated star cluster 
likely represents the nucleus of a captured galaxy undergoing the final
digestion stages of merging and assimilation with NGC~4424.
As the likely inner seed of a galaxy harvested from the nearby 
environment and consumed by NGC~4424, we refer to this star system as 
Nikhuli\footnote{Taken from the Sumi language, spoken by a 
  quarter of a million (formerly all tribal) people in Nagaland, the name
  relates to the Tuluni, {\it aka} Anni, festival. (Many Naga tribes are
  known to have been headhunters into the 19th and 20th century.)  
\label{foot:Nikhuli}}, which is a name 
relating to a festive period of celebrating and wishing for a rich harvest. 

The central black hole of Nikhuli, should one exist,  might be smaller
than the above prediction, and thus also of intermediate-mass ($10^2$--$10^5\,M_{\odot}$).  
\citet{2020MNRAS.492.3263G} has found that the $M_{\rm bh}$--$M_{\rm nsc}$
relation for nuclear star clusters (NSCs) also applies to ultracompact dwarf (UCD)
galaxies, thought to be the tidally stripped nuclei of galaxies
\citep{1988IAUS..126..603Z, 1993ASPC...48..608F, 2016ApJ...824...10F}.  
Nikhuli likely represents such a system. 
From a visual inspection of
the `residual image' of NGC~4424, it is hard to establish how much of the
infalling nuclear star cluster may have been stripped away, or instead, how much of
the infalling galaxy still remains around the infalling nuclear star cluster. 
Nonetheless, we can employ equation~7 from 
\citet{2020MNRAS.492.3263G}, which relates the stellar mass of nuclear star clusters
to their central black hole mass.  The relation can be written as 
\begin{eqnarray}
\log(M_{\rm bh}/{\rm M}_{\odot}) & = & 
(2.62\pm0.42)\log(M_{\rm nc}/[10^{7.83}\,{\rm M}_{\odot}]) \nonumber  \\
 & &  + (8.22\pm0.20), \label{Eq-nsc}
\end{eqnarray}
with the uncertainty on the black hole mass given by the expression 
\begin{eqnarray} 
(\delta \log M_{\rm bh}/{\rm M}_{\odot})^2 & = &
 [ \log(M_{\rm nc}/[10^{7.83}\,{\rm M}_{\odot}]) ]^2(0.42)^2    \nonumber \\
 & & \hskip-40pt  + (0.20)^2 + (2.62)^2 [\delta \log M_{\rm nc} ]^2 + \epsilon^2. \nonumber
\label{EqMerr}
\end{eqnarray}
The term $\delta \log M_{\rm nc}$ is the uncertainty associated with the nuclear
star cluster's stellar mass, and $\epsilon$ is the intrinsic scatter in
the ($\log M_{\rm bh}$)-direction, and taken to be 1.31 dex
\citep{2020MNRAS.492.3263G}.

Assigning a factor of 2 uncertainty to the total 
stellar mass of the stretched star cluster, 
one obtains a (poorly constrained)\footnote{The heightened uncertainty is due
  to the steep slope of, and scatter about, the $M_{\rm bh}$--$M_{\rm nsc}$ relation.}  value of 
$\log(M_{\rm bh}/{\rm M}_{\odot})=4.83\pm1.63$, corresponding to 
$M_{\rm bh} = 7\times10^4\,M_{\odot}$ when expressed in linear units. 
For reference, due to the steep (2.62) slope of the 
$M_{\rm bh}$--$M_{\rm nsc}$ relation (Equation~\ref{Eq-nsc}), a factor of 2 reduction to the star
cluster mass results in a large 0.8 dex decrease to the predicted black hole
mass in the star cluster.
This black hole mass should not be confused with the above
predictions for a {\it central} massive black hole in NGC~4424. 

For the potential nuclear star cluster already at the center of NGC~4424, if it has been
experiencing continuous star formation for the past 0.5~Gyr 
\citep{2018A&A...620A.164B}, then the associated $M/L_H$ ratio 
could be as low as 0.13, or half that value if the star formation started
0.1~Gyr ago  \citep[][see their Figure~13]{2014A&A...561A.140B}. 
This uncertainty, coupled with the dust extinction at the center of the
galaxy, which is evidently not negligible 
in the $H$-band, makes it problematic to ascertain a reliable
mass for the potential nuclear star cluster at the center of NGC~4424.
We have, therefore, not attempted to estimate a stellar mass. 

From $L_{\rm X} \equiv L_{0.5-10\,{\rm keV}} = 7.09\times 10^{38}$ erg s$^{-1}$
(Table~\ref{Tab_Sum}), the Eddington ratio ($L_{\rm X}/L_{\rm Edd}$) of NGC~4424 X-3
can be 
calculated using $L_{\rm Edd} = 1.3\times10^{38}\, M_{\rm bh}/M_{\odot}$
erg s$^{-1}$.  Given the above
black hole mass estimate of $7\times10^4\,M_{\odot}$, the 
Eddington ratio is $8\times10^{-5}$, or roughly $10^{-4}$.  Such a value is
typical of those seen in the few active IMBH candidates at the centers of
early-type galaxies in the Virgo cluster \citep[][their Figure~8]{2019MNRAS.484..794G}. 
Alternatively, NGC~4424 X-3 could be a 5.5 solar mass black hole radiating at
the Eddington limit.  

\subsubsection{X-ray Binaries}

The likelihood of NGC~4424 X-3 being a compact stellar mass object, and XRB, 
such as an accreting neutron star or a stellar-mass black hole, is not high. 

High-mass X-ray binaries (HMXBs), in which the donor star has a high mass, obviously requires
a young donor star.  However, the morphology and red color of Nikhuli suggests that it 
consists of an older, evolved, stellar population, thereby disfavouring a
HMXB in Nikhuli.  We can estimate the probability of a field HMXB in NGC~4424 having a
chance alignment with Nikhuli by using the knowledge that the 
abundance of HMXBs scales with the galaxy star-formation rate 
\citep{1978ComAp...7..183S, 2001ApJ...559L..97G, 2003MNRAS.339..793G,
  2013ApJ...764...41F}. 
  \citet{2015A&A...579A.102B} report a global star formation rate of 0.32
  $M_\odot$ yr$^{-1}$ in NGC~4424, rescaled from 23~Mpc to our distance of
  16.4 Mpc, while \citet{2018A&A...620A.164B} revise this down further to 0.25
  $M_\odot$ yr$^{-1}$. This corresponds to a 
specific star-formation rate of $\log(\rm{sSFR}\,{\rm yr}^{-1}) \approx -10.4$,
based on $M_{*,gal} = -9.8\pm0.4$ \citep{2019MNRAS.484..814G}. 
For $\Gamma=1.7$, 
the X-ray point-source in Nikhuli has $L_{0.5-8\,{\rm keV}} \approx 6 \times 10^{38}$
erg s$^{-1}$. Now, from\citet{2019ApJS..243....3L}, one therefore 
 expects just two HMXBs with $L_{0.5-8\,{\rm keV}} \ge 6\times10^{38}$ erg s$^{-1}$
 across the galaxy's $\sim$9 square arcminutes of star-forming area, thereby
 making the arcsecond-alignment of NGC~4424 X-3, should it be a HMXB, 
with Nikhuli an extremely unlikely coincidence.

We can also explore the probability that NGC~4424 X-3 is a low-mass X-ray
binary (LMXB). 
The X-ray luminosity function of LMXBs scales with the stellar mass. 
From \citet{2019ApJS..243....3L}, one again expects around two LMXBs with 
$L_{0.5-8\,{\rm keV}} \ge 6\times10^{38}$ erg s$^{-1}$ per $10^{11}\,M_\odot$, 
or $0.7\times10^{-4}$ per $3.44\times10^6\,M_\odot$, which is the mass of
Nikhuli.  However, this expectation is based upon field star statistics.
The number of LMXBs in globular clusters is $\sim$10$^3$ times greater than in the field
\citep{2007ApJ...660.1246S,2013ApJ...764...98K,2020ApJS..248...31L}.
As a former nuclear star cluster, Nikhuli may thus have a $\sim$7\% chance of
being an LMXB, {\it if} globular clusters and nuclear star clusters have a similar
LMXB formation efficiency. This is the only realistic chance NGC~4424 X-3 has 
of being an XRB rather than a massive black hole. 
However, supermassive black holes and nuclear star clusters are known
to regularly coexist in galaxies with $10^6 < M_{\rm bh}/M_\odot < 10^7$
\citep[e.g.][and references therein]{2009MNRAS.397.2148G}, 
and thus the infall origin for Nikhuli {\it may} favor the
presence of an active massive black hole rather than an XRB.

\section{Infall timescale}

Lacking an outward-pointing, 
comet-like appearance, we suspect that this active\footnote{We use
  the term `active' to denote a star cluster with an active black hole; in
  this case, X-ray active.}  star cluster is probably not an ejected nucleus
from a gravitational wave recoil event 
\citep{2008MNRAS.390.1311B, 2008ApJ...686..829H, 2008ApJ...689L..89K,
  2021MNRAS.502.2682A, 2021MNRAS.503.1688H, 2021ApJ...913..102W}.
Although probably evident to many readers, we note that 
Nikhuli has a lopsided shape, unlike that of a (background) galaxy. 
The significant elongation of Nikhuli in the direction of the NGC 4424 galaxy
center favors an infall scenario. Conceivably, three additional knots to the
north may delineate the orbital path of the infalling galaxy.  Tidal 
stretching \citep{Roche:1850} 
of an accreted nuclear star cluster, and possibly the paltry remains of the accreted galaxy
still surrounding the nuclear star cluster, is evident.

There is an abundance of nucleated dwarf and nucleated spiral galaxies
\citep{2002AJ....123.1389B, 2003AJ....125.2936G}, and their
nuclear star cluster can contain a massive black hole 
\cite[e.g.][]{2009MNRAS.397.2148G}.  The accretion of these galaxies, coupled
with dynamical friction, can 
drive their nuclear star clusters toward the center of the accreting
galaxy. 
Massive structures, such as dense star clusters, can sink due to the braking force 
and thus orbital decay, arising from the dynamical friction with the
surrounding galaxy \citep{1943ApJ....97....1C, 1974SvA....18..180B,
  1975ApJ...196..407T, 2011MNRAS.416.1181I, 2014ApJ...785...51A,
  2014MNRAS.444.3738A, chen2021dynamical, morton2021gaseous}.  

Differing from the {\em
  naked}\footnote{The term `naked' is used to denote the absence of a
  surrounding star cluster.} black holes in the  merger simulations of
\citet{2021MNRAS.505.5129B} and \citet{2021MNRAS.508.1973M}, which tend not to sink to
the center of the final galaxy but remain off-center, the {\em shrouded} black hole in Nikhuli 
may have more success \citep{2019MNRAS.486..101P, 2021MNRAS.508.1973M}. 
\citet{2010MNRAS.401.2753B} have shown that star clusters with masses greater
than $2\times10^5\,M_\odot$, i.e., less than 10 times the mass of Nikhuli,
experience significant orbital decay within 1 Gyr. 
Other simulations have found that black holes can form close pairs in dwarf galaxies, 
which likely merge to build the foundation of the central supermassive black hole 
and produce gravitational wave signals detectable using the planned Laser
Interferometer Space Antenna  \citep[e.g.,][]{2019MNRAS.482.2913B}. 
In addition, \citet{2000ApJ...543..620O} report that globular cluster capture 
may explain some of the nucleated dwarf galaxies in the Virgo cluster, 
with the clusters reaching the core within a Hubble time. 

Following \citet[][their eq.~7.18]{1987gady.book.....B}, 
a crude estimate for the time, $t$, it would take Nikhuli (with NGC~4424 X-3) to sink to
the center of NGC~4424 can be made when 
assuming the inner region of NGC~4424 is dominated by stars with a roughly
homogeneous, isothermal distribution.  Their dynamical friction timescale is given by 
\begin{equation} 
t = \frac{264}{\ln \Lambda} 
\left( \frac{10^6\, {\rm M}_\odot}{M} \right) 
\left( \frac{r}{2\, {\rm kpc}} \right)^2 
\left( \frac{v_c}{250\, {\rm km\, s}^{-1}} \right) {\rm Gyr}.
\end{equation}
While \citet{1999ApJ...523..566C} adopted $\log \Lambda = 10$ for 
Coulomb's logarithm, we use $\ln \Lambda = 10$, which is 2.3 times smaller.
Substituting in $M=3.44\times10^6$ M$_\odot$ for Nikhuli, along with 
$r=400$ pc and 
$v_c = \sqrt{2}\sigma$, where $\sigma=57$ km s$^{-1}$, 
one obtains a dynamical friction time of $\sim$100 Myr. 
If one assumes that Nikhuli is at a non-projected distance of 1 kpc from the
galaxy center, then one obtains a 
timescale of $\sim$0.64 Gyr.  Such timescales mesh well with those seen in 
\citet[][see their Figure~2]{2001ApJ...552..572L}, who used a value of $\ln
\Lambda \approx 7.3$.  If we adopt the value $\log \Lambda = 2$ found by
\citet{1999MNRAS.304..254V} for bulgeless, exponential disks in dark matter halos
with NFW \citep{1996ApJ...462..563N} profiles, then the above timescales increase to
$\sim$0.5 and $\sim$3.2 Gyr.  However, as \citet{2004ApJ...605L..13M} note, 
disk perturbations, which can include spiral arms and bars, can result in both positive and
negative torques on an infalling satellite.  
The bar-like feature in NGC 4424 may facilitate the 
inspiral of the Nikhuli/black hole system \citep{bortolas2021role}.  
Whether or not Nikhuli makes it all the way to the center of NGC~4424, 
and how long that may take, will also depend on the time spent in the disk plane and
requires modeling beyond the scope of this paper. 

In low-mass spheroids with low S\'ersic indices, and thus a shallow
gradient to their central gravitational potential
\cite[e.g.,][]{2005MNRAS.362..197T, 2007MNRAS.377..855T}, star clusters can
wander about their galaxy's center, 
typically displaced by $\sim$100 parsecs \citep{2000A&A...359..447B}.
Massive black holes can also be offset 
\citep{2020MNRAS.495L..12B, 2021MNRAS.503.6098R}.
A similar situation occurs in the partially depleted cores of massive
spheroids, where infalling perturbers can stall outside of the cores with
shallow density profiles 
\citep{2006MNRAS.368.1073G, 2006MNRAS.373.1451R, 2009MNRAS.397..709I, 
2015MNRAS.454.3778P, 2016ApJ...829...81B, 2021ApJ...912...43B}. 
However, 
Figure~\ref{Fig_4424_resid} does not suggest that Nikhuli has stalled, and as
such we can not conclude that Nikhuli is destined to become an eternally wandering
black hole.  We can, however, derive an additional estimate of the dynamical
friction time scale for a shallow potential. 

Assuming that Nikhuli is now orbiting within the pseudobulge of NGC~4424, and
assuming this pseudobulge dominates the inner gravitational potential, 
we can use the S\'ersic function which was found to describe this structure
(Figure~\ref{Fig-Prof}) to estimate the infall time for Nikhuli. 
Building on \citet[][their equation~21]{2014ApJ...785...51A}, 
\citet[][their equation~8]{2015ApJ...806..220A} provide an (orbital
eccentricity)-dependent variant from which the following 
expression was obtained.  The equation depends on the inner profile slope and 
yields the time an infalling star cluster of mass $M_{\rm sc}$ at
radius $r$ will take to get to the center of a bulge/halo with mass $M_{\rm b}$,
scale radius $r_s$, and negative logarithmic slope of the inner density profile
$\gamma_0$. 
\begin{equation}
t ({\rm Myr}) = 0.3\sqrt\frac{(r_s/{\rm kpc})^3}{(M_{\rm b}/10^{11}\, {\rm M}_\odot)} g(e,\gamma_0)\left(\frac{M_{\rm
    b}}{M_{\rm sc}} \right)^{0.67} \left( \frac{r}{r_s} \right)^{1.76}.
\label{Eq_Arc}
\end{equation}

From our fitted S\'ersic function, we derived a mass for the
pseudobulge\footnote{Using the pseudobulge mass in the $M_{\rm bh}$--$M_{\rm
    bulge}$ relation for late-type galaxies \citep{2019ApJ...873...85D,
    2019ApJ...876..155S} gives a predicted black hole mass of $\log(M_{\rm
    bh}/{\rm M}_\odot)=3.8$.  However, the unrelaxed nature of this pseudobulge
  makes this a dubious prediction.} such that $\log(M_{\rm b}/{\rm M}_\odot) =
8.48\pm 0.34$, assuming $M/L_{F160W} = 0.5$.  This is notably greater than the
mass of Nikhuli with $M_{\rm sc} = (3.44\pm0.93)\times10^6\,{\rm M}_{\odot}$,
which we take to be at a deprojected radius $r=5\arcsec$ (395~pc) from the
center of NGC~4424.  \citet[][their equation~23]{2006AJ....132.2701G} provide
an equation for the negative logarithmic slope, $\gamma_0$, of the model from
\citet{1997A&A...321..111P}, which uses the S\'ersic model parameters to
provide an expression which approximates the deprojected
\citet{1963BAAA....6...41S} $R^{1/n}$ model. For small radii, $\gamma_0$
depends solely on $n$. We found the pseudobulge in NGC~4424 is well-described
with $n = 0.47 \pm 0.08$, for which we obtain $\gamma_0 \approx 0$.
\citet[][their equation~5]{2015ApJ...806..220A} provide an expression to
derive the appropriate scale radius $r_s$ from the projected `effective
half-mass radius' of the bulge/halo --- taken here to be the `effective
half-light radius' of our equivalent-axis ({\it aka} geometric-mean axis)
model for the pseudobulge.  With $R_{\rm e,eq} = 3\farcs 87\pm 0\farcs39$ from
our S\'ersic model fit to the pseudobulge, we have that $r_s \approx
0.35R_{\rm e,eq} = 1\farcs35$, or $\approx$106 pc.  Finally, for an
eccentricity $e=0$ and $\gamma_0 = 0$, the $g(e,\gamma_0)$ term above, from
\citet[][their equation~10]{2015ApJ...806..220A}, equals 5.83.  Inserting the
above values into equation~\ref{Eq_Arc}, one obtains a dynamical friction time
of 220 Myr.

Alternatively, we can approximate the total (bulge $+$ bar $+$ disk) light
over $\sim$1 to $\sim$8 arcseconds in Figure~\ref{Fig-Prof} with an $n=1$ model having
a half-light radius of 8$\farcs$4.  The stellar mass of this model, within this
half-light radius, is approximately $0.8\times 10^9$ M$_\odot$, with an
extrapolation of the profile to 
infinity doubling this mass to $1.6\times 10^9$ M$_\odot$. 
Under the approximation of a spherical
system, this model corresponds to $\gamma_0 = 0.44$, $r_s = 3\farcs5$ (277 pc), and 
$g(e,\gamma_0) \approx 5.2$, and it yields a similar infall time of 206 Myr. 
Increasing the assumed distance of Nikhuli from the galaxy center by a factor
of 3 lengthens the infall time to $\sim$1.4 
Gyr, while assuming an eccentricity of 0.5 will reduce the times by a factor
of 0.65, giving $t=0.9$ Myr in this instance. 
The star cluster(s) seen within the inner arcsecond may be a testament to these 
dynamical times.

\section{Discussion}\label{Sec_Disc}

The phenomenon of an infalling galaxy undergoing assimilation has been 
offered to explain the offset IMBH in ESO~243-49 \citep{2009Natur.460...73F,
2010ApJ...721L.148B, 
2012MNRAS.423.1309M, 2017A&A...602A.103W, 2021MNRAS.505.5129B, 2021MNRAS.503.6098R} 
and the ULX sources likely associated with a massive black hole
in galaxies such as IC~4320 \citep{2012MNRAS.423.1154S}, 
NGC~5252 \citep{2015ApJ...814....8K, 2020MNRAS.493L..76K}, NGC~2276-3c
\citep{2015MNRAS.448.1893M} and others. 
One such interesting example can be seen in the Virgo
cluster galaxy NGC~4651 \citep{2010AJ....140..962M, 2014MNRAS.442.3544F,
  2018A&A...614A.143M}, where the elongation process associated with infall 
has produced a tidal stream \citep{2016ASSL..420....1N} which
is immediately evident in optical images. 
An earlier stage of such 
shredding, of an infalling dwarf galaxy, has been reported in a galaxy 
at a redshift $z=0.145$ by \citet{2003Sci...301.1217F}.  See also UGC~10214
\citep{1959VV....C......0V, 2011A&A...536A..66M}, NGC~7714 
\citep{2004A&A...422..915S}, and M32 
giving itself to M31 \citep{1964ApJ...139.1045A,
  2002ApJ...568L..13G, 2006ApJ...638L..87G}. 
Various stages of the interaction process are evident in the catalog of 
\citet{1977A&AS...28....1V}, and the  
scenario in which a galaxy is threshed and reduced to its dense core 
has been developed by 
 \citep{2003MNRAS.346L..11B}. 

In and around our Galaxy, we know that the Sagittarius dwarf 
galaxy is undergoing disruption \citep{1994Natur.370..194I, 2015MNRAS.446.3110K}, 
the Cetus and Indus stellar streams are the remnants of accreted dwarf
galaxies \cite{2020ApJ...905..100C, 2021ApJ...915..103H}, 
and the Gaia-Enceladus-Sausage was acquired long ago 
\citep{2018MNRAS.478..611B, 2018Natur.563...85H}.  Furthermore, 
\citet{2019NatAs...3..667I} revealed the impressive tidal stream associated with
$\omega$ Centauri, possibly the pared-core of an accreted dwarf galaxy 
but currently lacking confirmation of a central black hole 
\citep{2019MNRAS.482.4713Z, 2019MNRAS.488.5340B}.
Additional globular cluster streams, evidence of accretion and possibly 
remnant galaxy nuclei in the Milky Way are known
\citep[e.g.][]{2021ApJ...909L..26B, 2021AJ....162...42W, 
  2021MNRAS.500.2514P, 2020ApJ...902...89T, ferguson2021delveing,
  2021A&A...646A.176P}. 
\citet{2020ApJ...898L..37Y} and \citet{2021ApJ...920...51M} discovered and
investigated the low-mass stream 
LMS-1, revealing the remnants of a low-mass galaxy now just 10 to 20 kpc from our Galaxy's
center, making it the closest known stream to the Galaxy's center thus far.
They conclude that the ``globular cluster'' NGC~5024 or 
NGC~5053 is likely the former nuclear star cluster of this low-mass galaxy, akin to 
NGC~5824 possibly being the remnant nuclear star cluster of the dwarf
progenitor to the Cetus stream \citep{2020ApJ...905..100C}.
The slightly more massive (than globular cluster) 
UCD galaxies \citep[e.g.][]{2017ApJ...839...72A} 
have also been caught in the act of stripping.  They are considered to be
the former nuclei of galaxies
\citep[e.g., VUCD3:][]{2015ApJ...812L...2L} and are known to contain massive black
holes \citep[][and references therein]{2020MNRAS.492.3263G}.

Globular clusters are known to
contain XRBs \citep[e.g.][]{1992PASP..104..981H, 2003ApJ...594..798B,
  2003ApJ...586..814M}, and it has been speculated that their capture may
contribute to the LMXB population in galaxies \citep{2020ApJS..248...31L}. 
With a stellar mass of $\sim$3$\times10^6\,M_\odot$, Nikhuli is more massive
than most globular clusters. 
Furthermore, the width of the Nikhuli stream is some 35 pc across. 
As such, we are probably not seeing a captured globular cluster---which have half
light radii typically less than 4~pc \citep{2005ApJ...634.1002J}---but 
more likely the nucleus of a stripped galaxy. 
Conceivably, the color or metallicity of Nikhuli might shed light on the full
stellar mass
of the progenitor via the color-magnitude or mass-metallicity relation for
dwarf galaxies.  However, 
given that the remnant nuclear star cluster light will now dominate the flux and color, 
this would yield questionable results. 

Although Nikhuli is unlikely to be a captured globular cluster, if nuclear
star clusters have a similar LMXB formation efficiency as globular clusters,
then Nikhuli might contain a LMXB.  It would therefore be desirable to observe
Doppler-broadened, optical emission lines coming from NGC~4424 X-3 to verify
the existence of the potential massive black hole.  Such confirmation of a
massive black hole mass might be possible if (with sufficiently high spatial
resolution in order to reduce the stellar noise) broad optical emission lines
are observed from NGC~4424 X-3.  An alternative approach to estimate the black
hole mass could come from the `fundamental plane of black hole activity'.  As
given by \citet{2012MNRAS.419..267P}, it predicts a radio luminosity
$\nu\,L_{\nu}$(5~GHz) of $10^{33.9}$ erg s$^{-1}$ for $M_{\rm bh}=0.7\times
10^5\,M_\odot$ and $L_{0.5-10\,{\rm keV}} = 0.7\times 10^{39}$ erg
s$^{-1}$.  At a distance of 16.4~Mpc, this corresponds to 5 $\mu$Jy, 
tantalizingly within the reach of the Karl G.\ Jansky Very Large Array (VLA) and the
Australia Telescope Compact Array (ATCA), which have probed  
down to root mean square (rms) noise levels of $\sim$1--2 $\mu$Jy beam$^{-1}$
 \citep{2012ApJ...750L..27S, 2018ApJ...862...16T}.

An incoming black hole of equal mass to that already (presumed to be) at the
center of NGC~4424 might imply a major merger between two spiral galaxies.
However, such a scenario is unlikely to result in a post-merger galaxy looking
like a spiral galaxy.  
On the other hand, early-type galaxies have smaller stellar masses for a given
black hole mass than late-type galaxies 
\citep{2019ApJ...876..155S}.  As such, a minor merger {\it can} bring in an
equal mass black hole.  This suggestion, which we think is presented here for
the first time, may also explain why the extreme ULX and HLX off-center
sources found by \citet{2012MNRAS.423.1154S} in eight nearby ($<$100 Mpc)
galaxies are not in galaxies displaying signs of a recent major merger.

A minor merger in NGC~4424 also meshes with the notion of an unequal (stellar
mass) merger reported by \citet{2018A&A...620A.164B}.  From Figure~11 in
\citet{2019ApJ...876..155S}, one can see that an early-type galaxy with a
black mass of $10^{4.8}\,M_{\odot}$ (see section~\ref{Sec-BH-Nik}) will have
a stellar mass of $\sim0.6\times10^9\,M_{\odot}$, which is just 6\% of the
stellar mass of NGC~4424 (see section~\ref{Sec_Opt}).  Suppose such an early-type
galaxy is what fell into NGC~4424.  In that case, (ignoring gaseous processes), the
coalescence of NGC 4424's initial and incoming black hole will double the
central black hole mass, with the stellar mass increasing by only $\sim$6\%.
We speculate that such minor mergers may help maintain the steepness of the
non-linear, near-cubic $M_{\rm bh}$--$M_{\rm *,gal}$ relation for spiral
galaxies \citep{2016ApJ...817...21S, 2018ApJ...869..113D}.  Furthermore,
compact spheroids and the bulges of early-type galaxies formed early in the
Universe \citep[][and references therein]{2015ApJ...804...32G} and were 
likely the formation sites of the first massive black holes.  Some of these
galaxies (and their black holes) may have subsequently been sewn (and sown)
into late-type spiral galaxies.

Given the similarity of the predicted (central and infalling) black hole
masses in NGC~4424, and the absence of an X-ray point source at the (very)
center of NGC~4424, it might, but need not, be the case that NGC~4424
currently has no central black hole.  The delivery of NGC~4424 X-3 via a
low-mass early-type galaxy might demonstrate how some apparently bulgeless late-type
galaxies, like, for example, M33 
\citep{2001Sci...293.1116M}, could eventually be both seeded with massive black
holes and commence building their bulge.  
The median bulge-to-total flux ratios in Sc and Sd
galaxies are only 8\% \citep{2008MNRAS.388.1708G}. 
The bulge growth from minor mergers might be slow at 
first, and initially go undetected while the bulge-to-total stellar mass ratio
is small or ill-defined, leading to `bulgeless galaxies' with massive black
holes, such as NGC~4395, NGC~2748 and NGC~6926 \citep{2019ApJ...873...85D}.
Indeed, \citet{2006AJ....131..747C} has suggested that NGC~4424 will become a
lenticular galaxy with a small bulge over the next 3 Gyr.  The merger-induced
bulge growth arises not just from the delivery of new stellar material sewn 
into the galaxy, but also from the redistribution of gas toward the galaxy center, which
subsequently undergoes star formation.  

In a downsizing scenario, such minor mergers would represent the late-time seeding of massive
black holes into late-type spiral galaxies.  
Seeding by accretion may, of course, occur over a range of
redshifts.  This is different to high-$z$ quasars getting a kick start in life
from a `black hole seed' that is an IMBH
\citep[e.g.,][]{2004PhRvD..70f4015D,2004MNRAS.354..292K}, 
unless the origin of massive black holes in dwarf early-type galaxies is the
same as high-$z$ quasars.  While the old nuclear star clusters of five
low-mass dwarf early-type galaxies \citep{2011MNRAS.413.1764P} may support this, those clusters 
could represent globular clusters captured through dynamical friction. 
Furthermore, it is not yet clear that today's 
dwarf early-type galaxies, a source of minor mergers for spiral galaxies, 
ubiquitously contain a massive black hole. 
Although \citet{2021Graham} report that
the detection of an X-ray point-source at the center of low-mass 
late-type galaxies is $\sim$40\%, while it is just $\sim$10\% in low-mass early-type
galaxies, they suggest that this may reflect the availability of cold gas for 
igniting the candidate AGN in these galaxies, rather than representing a measure of the
massive black hole occupation fraction.  Further work is required to establish
the true massive black hole occupation fraction in low-mass galaxies.

In addition to a 
`classical' (i.e., merger built) bulge, an incoming perturber might cause a thin
in-plane bar to experience an 
instability leading to the creation of an X/(peanut shell)-shaped
`pseudobulge' structure \citep{1990A&A...233...82C, 2005MNRAS.358.1477A, 
2016MNRAS.459.1276C}.  Such a feature may be apparent in NGC~4424 \citep[][see
  their Figure~2]{2006AJ....131..747C}, where, interestingly, there are also
signs that this structure forms 
the inner half of a figure-of-eight pattern, such as that more clearly seen in NGC~4429
\citep[][their Figure~6]{1996AJ....111..174F, 2011A&A...532A..74B,
  2011MNRAS.413..813C}.  
In passing, we speculate that
NGC~4429 may, therefore, represent a more evolved state of NGC~4424.
NGC~4608 \citep[][]{2011MNRAS.413..813C} may offer a more face-on representation of NGC 4429.
If so, then the figure-of-eight pattern may arise from an X-shaped pseudobulge coupled with the
ansae/ring at the end of the original bar.

At $6\times 10^{10}\,M_{\odot}$ \citep{2015ApJ...806...96L}, the Milky Way has a stellar mass some
six times bigger than NGC~4424, and a clear X/P pseudobulge \citep[][and
  references therein]{2017MNRAS.471.3988C}.  This pseudobulge, which peaks at
roughly half the length of the bar, likely formed from 
perturbations to the bar due to by  minor fly-bys 
\citep{2021MNRAS.506...98K} or minor mergers which also contribute to the 
classical bulge \citep{2017PASA...34...41N, 2018A&A...618A.147Z}. 
If six minor merger events have brought in
six $10^5\,M_{\odot}$ black holes, one might (naively) expect a central black
hole with a minimum mass of $0.6\times 10^6\,M_{\odot}$, with gas fuelling
and stellar capture perhaps increasing this to the observed value of 
$4\times 10^6\,M_{\odot}$ \citep{2016ApJ...830...17B}.  The zoom-in hydrodynamical
simulation of \citep{2016MNRAS.459.2603B} supports such growth by infall and
black hole merging.  It would also explain the tentative
suggestions of IMBHs infalling/wandering in the Milky Way 
\citep[][and references therein]{2020ApJ...890..167T}.

\section{Summary}

We have discovered, in {\it HST} imaging, a red, elongated star cluster located 
$\sim$400~pc ($\sim$5$\arcsec$),  in projection, from the center of the
post-merger galaxy NGC~4424.  This may be the remnants of the 
infalling galaxy responsible for the known, past minor-merger event that
disturbed NGC~4424.  The star cluster, referred to as Nikhuli, is stretched in the direction of
NGC~4424's center, and could be the nuclear star cluster of the
captured galaxy.  

Based on the $3\times10^6\,M_\odot$ stellar mass of Nikhuli, we used the
$M_{\rm bh}$--$M_{\rm nc}$ scaling relation to predict that it may harbour an
intermediate-mass black hole of $7\times10^4\,M_\odot$, although this
estimate has a considerable 1.6 dex of uncertainty.  Interestingly, Nikhuli
corresponds spatially with a {\it CXO} X-ray point-source having $L_{\rm X} \equiv
L_{0.5-10\,{\rm keV}} = 0.7\times 10^{39}$ erg s$^{-1}$ for a photon index
$\Gamma=1.7$.  While this source may be a super-Eddington stellar-mass black
hole, we argue that it is more likely to be a sub-Eddington massive black
hole.

Based on the stellar mass and velocity dispersion of NGC~4424, it is expected
to possess an intermediate-mass black hole with $\log M_{\rm bh}/{\rm M}_\odot=4.9\pm0.6$. 
While the X-ray point-source in Nikhuli is highly unlikely to be a high-mass
X-ray binary, we estimate that there is a 7\% chance that it may be a low-mass
X-ray binary {\it if} the formation efficiency of LMXBs in nuclear star
clusters is 1000 times greater than that of a galaxy's field stars. 
On the other hand, given that nuclear star clusters and massive black holes
are known to regularly coexist, Nikhuli may be the delivery vehicle for
seeding a, or at least helping to grow the, massive black hole in NGC~4424. 
At the same time, Nikhuli's now disrupted host galaxy may have been sewn into
the fabric of NGC~4424, contributing to the build up of the stellar halo or a
classical bulge depending on how far the stars penetrate. 

We plan to search for more such 
hidden structures, like Nikhuli, in 74 other Virgo cluster spiral galaxies
imaged by {\it CXO} \citep{Soria2021}.  To date, a lot of the tidal stream
discoveries have occurred in the halos of galaxies.  Our galaxy capture-and-subtract process
will hopefully enable one to better probe the inner regions.  In particular, we intend
to cross-match any new discoveries with the off-center ULXs identified by 
\citep{Soria2021}.  This may 
provide a statistical argument for how important minor merging could be as
either a black hole seeding or growth mechanism for late-type spiral galaxies.

\begin{acknowledgments}

This research was supported under the Australian Research
  Council's funding scheme DP17012923 and 
is based upon work supported by Tamkeen under the NYU Abu Dhabi
 Research Institute grant CAP$^3$. 
RS warmly thanks Curtin University for their hospitality during the planning stage of
this project. 
Support for this work was provided by the National Aeronautics and Space
 Administration through Chandra Award Number LP18620568 issued by the Chandra
 X-ray Center, which is operated by the Smithsonian Astrophysical Observatory
 for and on behalf of the National Aeronautics Space Administration under
 contract NAS8-03060.
Data underlying this article are available in the Chandra Data Archive
 (CDA: \url{https://cxc.harvard.edu/cda/}), 
 and the Next Generation Virgo Cluster Survey website
 (\url{https://www.cfht.hawaii.edu/Science/NGVS/}). 
Based on observations made with the NASA/ESA Hubble Space Telescope, and
 obtained from the Hubble Legacy Archive, which is a collaboration between the
 Space Telescope Science Institute (STScI/NASA), the Space Telescope European
 Coordinating Facility (ST-ECF/ESA) and the Canadian Astronomy Data Center (CADC/NRC/CSA).
This research has made use of NASA’s Astrophysics Data System (ADS) 
 Bibliographic Services and of the NASA/IPAC Extragalactic Database (NED), 
 which is operated by the Jet Propulsion Laboratory, California Institute
 of Technology, under contract with NASA. 

\end{acknowledgments}

\software{
{\sc IRAF} \citep{1986SPIE..627..733T,1993ASPC...52..173T}, 
{\it Isofit/Cmodel} \citep{2015ApJ...810..120C}, 
{\sc Profiler} \citep{2016PASA...33...62C}, 
{\sc xspec} \citep[v12.11.0:][]{1996ASPC..101...17A}, 
{\sc ciao} \citep[v4.12:][]{2006SPIE.6270E..1VF}, 
{\sc SAS} \citep[v17.0.0:][]{2004ASPC..314..759G}, 
{\sc ds9} \citep{2003ASPC..295..489J}. 
}

\bibliographystyle{aasjournal}
\bibliography{Paper-N4424}

\end{document}